\documentclass[11pt]{article}
\usepackage{amsmath,amsfonts}
\def \beq  {\begin{equation}}
\def \eeq  {\end{equation}}
\def \beqar {\begin{eqnarray}}
\def \eeqar {\end{eqnarray}}
\def\sqr#1#2{{\vcenter{\vbox{\hrule height.#2pt
\hbox{\vrule width.#2pt height#1pt \kern#1pt
\vrule width.#2pt}\hrule height.#2pt}}}}

\def\la {{\langle}}
\def\ra {{\rangle}}

\def\Tr {{\rm Tr}}
\def \tr {{\rm tr}}

\def\del {\partial}

\def\l {\lambda}

\def\D {{\cal D}}
\def\bz {{\bar{z}}}

\def\A {{\cal A}}

\def \Or {{\cal O}}

\def\half{\textstyle{1\over 2}}

\def \APNY {{\it Ann. Phys. NY}}
\def \CMP {{\it Commun. Math. Phys.}}
\def \PRL {{\it Phys. Rev. Lett.}}
\def \PLB {{\it Phys. Lett.} B}
\def \NPBProc {{{\it Nucl. Phys.} B ({\it Proc. Suppl.})}}
\def \NPB {{\it Nucl. Phys.}  B}
\def \RMP {{\it Rev. Mod. Phys.}}
\def \JGP {{\it J. Geom. Phys.}}
\def \JPA {{\it J. Phys. A: Math. Gen.}}
\def \CQG {{\it Class. Quant. Grav.}}
\def \MPLA {{\it Mod. Phys. Lett.} A}
\def \IJMPA {{\it Int. J. Mod. Phys.} A}
\def \JHEP {{\it JHEP}}
\def \PRB {{\it Phys. Rev.} B}
\def \PRD {{\it Phys. Rev.} D}
\def \JMP {{\it J. Math. Phys.}}
\def \GRG{{\it Gen. Rel. Grav.}}
\begin{document}
%%%%%%%%%%%%%%%%%%%%%%%%%%%%%%%%%%%%%%%%%%%%%%%
%\fontfamily{pnb}\fontsize{12pt}{16pt}\selectfont
%\fontfamily{pzc}\fontsize{14pt}{16pt}\selectfont
%\fontfamily{pbk}\fontsize{12pt}{16pt}\selectfont
\fontfamily{cmr}\fontsize{11pt}{18pt}\selectfont
%\fontfamily{phv}\fontshape{ro}\fontsize{11pt}{14pt}\selectfont
%\fontfamily{ptm}\fontseries{m}\fontshape{r}\fontsize{12pt}{16pt}\selectfont
%\fontfamily{pnc}\fontseries{m}\fontshape{r}\fontsize{11pt}{16pt}\selectfont
%\fontfamily{ppl}\fontseries{m}\fontshape{r}\fontsize{11pt}{14pt}\selectfont
%\usefont{T1}{phv}{m}{it}
%%%%%%%%%%%%%%%%%%%%%%%%%%%%%%%%%%%%%%%%%%%%%%%
%%%%%%%%%%%%%%%%%%%%%%%%%%%%%%%%%%%%%%%%%%%%%%%
\begin{titlepage}
\null\vspace{-62pt} \pagestyle{empty}
\begin{center}
\rightline{CCNY-HEP-06/8}
\rightline{June 2006}
\vspace{1truein} {\Large\bfseries
Quantum Hall Effect in Higher Dimensions, Matrix Models}\\
\vskip .2in
{\Large\bfseries  and Fuzzy Geometry}\\
\vskip .2in\noindent

%%%%%%%%%%%%%%%%%%%%%%%%%%%%%%%%%%%%%%%%%%%%%%%%%
\vspace{.1in}
{\bf\large D. KARABALI$^1$ and V. P. NAIR$^2$}\\
\vspace{.15in}{\itshape $^1$Department of Physics and Astronomy\\
Lehman College of the CUNY\\
Bronx, NY 10468}\\
E-mail: 
\fontfamily{cmtt}\fontsize{11pt}{15pt}\selectfont {dimitra.karabali@lehman.cuny.edu}
\vskip .1in
\fontfamily{cmr}\fontsize{11pt}{15pt}\selectfont 
{\itshape $^2$Physics Department\\
City College of the CUNY\\  
New York, NY 10031}\\
E-mail:
\fontfamily{cmtt}\fontsize{11pt}{15pt}\selectfont {vpn@sci.ccny.cuny.edu}

\vspace{.4in}
%\vspace{1.5in}%\vspace{0.3in}
%%%%%%%%%%%%%%%%%%%%%%%%%%%%%%%%%%%%%%%%%%%%%%%%%%%%%%%%%%%%
\centerline{\large\bf Abstract}
\end{center}
We give a brief review of quantum Hall effect in higher dimensions and its relation to fuzzy spaces.  For a quantum Hall system, the lowest Landau level dynamics is given by a one-dimensional matrix action whose large $N$ limit produces an effective action describing the gauge interactions of a higher dimensional quantum Hall droplet. The bulk action is a Chern-Simons type term whose anomaly is exactly cancelled by the boundary action given in terms of a chiral, gauged  Wess-Zumino-Witten theory suitably generalized to higher dimensions.
We argue that the gauge fields in the  Chern-Simons action can be understood as parametrizing the different ways in which the large $N$ limit of the matrix theory is taken. The possible relevance of these ideas to fuzzy gravity is explained. Other applications are also briefly discussed.
\end{titlepage}

%%%%%%%%%%%%%%%%%%%%%%%%%%%%%%%%%%%%%%%%%%%%%%%
\section{\bfseries Introduction}

It is well known that when the number of elementary quanta involved in any process
is very large, quantum dynamics can be approximated by classical dynamics; this
is the celebrated correspondence principle.  
The classical phase space takes over the role of the quantum Hilbert space. Quantum observables, which are linear hermitian operators on the Hilbert space, can be approximated by functions on the classical phase space.
Properties of functions on the phase space can be obtained as limits of properties of operators on the Hilbert space. 
Keeping this idea of correspondence in mind, the
general structure of a quantum field theory, describing gauge and matter fields, may be formulated 
as follows. We have an ambient spacetime differential manifold ${\cal M}$.
Fields are functions (or sections of an appropriate bundle) on ${\cal M}$.
They are also operators on the quantum Hilbert space of matter ${\cal H}_m$, and obey quantum conditions such as commutation rules, characterized by the deformation
parameter $\hbar$. At finite $\hbar$, we have the quantum field theory; as $\hbar \rightarrow 0$, we can approximate the physics by a classical field theory.

The general correspondence principle, however, suggests a further extension of this idea, and a new paradigm for physical theories. The spacetime manifold ${\cal M}$
itself may be viewed as an approximate method of description, obtained as the limit of some discrete Hilbert space ${\cal H}_s$. Thus, instead of functions on ${\cal M}$, physical fields
are operators on ${\cal H}_s$. They are also operators on the Hilbert space ${\cal H}_m$
of the theory. A new deformation parameter $\theta$, relevant to ${\cal H}_s$,
 may be introduced, so that,
as $\theta \rightarrow 0$, we can approximate the theory in terms of functions on a smooth manifold ${\cal M}$. Thus the usual quantum field theories are recovered in this limit.
(A further limit, $\hbar \rightarrow 0$, would take us to the classical field theory.)
In this formulation, fields are operators on ${\cal H}_s \otimes {\cal H}_m$,
or we may view them as matrix-valued quantum operators, the matrices being of dimension
$dim ({\cal H}_s )$. Field theories can thus be regarded as limits of matrix models.

The mathematical structure that is relevant here is that of fuzzy geometry, or,
more generally, noncommutative geometry \cite{bal1}. A fuzzy space is defined by a sequence of
triples 
$({\mathcal H}_N , Mat_N , \Delta_N )$,
where $Mat_N$ is the matrix algebra of $(N\times N)$-matrices which act  
on the
$N$-dimensional Hilbert space ${\mathcal H}_N$, and $\Delta_N$ is a  
matrix
version of the Laplace operator.
The matrices are taken to have an inner product given by, say,
$\la A,B\ra = {1\over N} \Tr (A^\dagger B)$, for arbitrary matrices $A, B$.
In the large $N$ limit, a matrix may be approximated by a function on some smooth manifold
${\cal M}$, the latter being a phase space corresponding to the Hilbert space
${\cal H}_N$. In this case, the deformation parameter  $\theta$ is a function of $N$, with $\theta \rightarrow 0$ as $N \rightarrow \infty$. At finite $N$,
we have the noncommutative algebra 
$Mat_N$, but this tends to the commutative
algebra of functions on
the smooth manifold ${\cal M}$ as $N\rightarrow\infty$. The Laplacian
$\Delta_N$ is used to define the metric and related geometrical
properties of the manifold
${\cal M}$. For example, information about the dimension of ${\cal M}$ is contained in
the rate of growth of the degeneracy of eigenvalues of $\Delta_N$.

Clearly the idea of formulating field theories as matrix models on a fuzzy space
is very appealing for a number of reasons.
The matrix formulation gives a discretization of the field theory and therefore, at the very least, 
we get a regularization of the theory with a finite number of modes.
This is analogous to the lattice regularization, but, in general, it is possible to
preserve more symmetries in a fuzzification than in latticization \cite{bal2}.
Secondly, and perhaps most importantly, space, or spacetime, is being viewed as an approximation to a Hilbert space ${\cal H}_s$. Thus the dynamics of spacetime geometry, in other words, gravity, can be naturally described as dynamics on the 
Hilbert space ${\cal H}_s$.
The fact that the number of modes would be finite in a fuzzy formulation will ensure that we
have a mathematically well defined formulation of gravity.

It is worth recalling at this stage that fuzzy geometry is part of the more general framework of noncommutative geometry. 
Noncommutative geometry is a generalization of ordinary geometry,
based on the following result.
The algebra of complex-valued square-integrable
functions on a manifold ${\cal M}$, with
pointwise
multiplication as the algebraic operation, is a commutative $C^*$-algebra. 
This $C^*$-algebra incorporates many of the
geometrical properties of the manifold ${\cal M}$. 
Conversely, any commutative
$C^*$-algebra can be represented by the algebra of functions on an
appropriate space ${\cal M}$, with the geometrical properties of
${\cal M}$ being images of corresponding algebraic properties
of the $C^*$-algebra.
This result allows a change of point of view: we may take the algebra as the fundamental
concept, the geometry being derived from it.
The generalization is then to consider a noncommutative
$C^*$-algebra; it may be taken as
the analog of an ``algebra of  
functions''
on some noncommutative space. The mathematical properties of this
noncommutative space are then implicitly defined by
the properties of the algebra. This is  
the
basic idea of noncommutative geometry \cite{connes1}-\cite{DN}.

More specifically, noncommutative geometry is given as a spectral triple
$(\A, {\mathcal H}, \D )$, where $\A$ is a noncommutative algebra with  
an
involution,
${\mathcal H}$ is a Hilbert space on which we can realize the algebra  
$\A$
as bounded operators and
$\D$ is a special operator which will characterize the geometry.
In terms of such a spectral triple,
the analog of differential calculus
on a manifold can be constructed. For the special case when
${\mathcal H}$ is the space of square-integrable spinor functions on a
manifold ${\cal M}$
(technically, sections of the irreducible spinor bundle), $\A$ is the
algebra of complex-valued
smooth functions on ${\cal M}$, and $\D$ is the Dirac operator
on ${\cal M}$ (for a particular metric and the Levi-Civita spin connection), 
the differential calculus constructed from the algebra is the
standard differential calculus on ${\cal M}$.
Going back to matrices, it
is clear that the algebra of finite dimensional matrices $Mat_N$ can play the role
of ${\cal A}$ and, hence, fuzzy geometry is a special case of noncommutative geometry.
(The idea of using noncommutative geometry  for gravity was suggested many years ago by Connes and others \cite{connes1}- \cite{connes2}.) 

While fuzzy spaces can be viewed as a regulator with real physics being eventually recovered when 
$N\rightarrow \infty$, the idea of fuzzy geometry goes further.
One may regard the true physics as given by the theory at finite, but large $N$,
the smooth manifold limit being a convenient simplification
for calculations. After all, it is an elementary truism that, while we formulate
physical theories on continuous spaces, infinite dimensional Hilbert spaces, etc.,
we always deal with a finite set of measurements, or even a finite number of possibilities for measurements. Therefore, it is almost tautological that physical theories, at least for
the case of space being even dimensional, can be described by finite dimensional matrix models.

Indeed, matrix models have recently appeared in a number of different contexts in physics.
It was observed many years ago that one could use matrix models as a regularization of membrane theories \cite{dewit}. By now this is well understood and matrices have become a standard technique for analysis of branes of different dimensions. Matrix models descriptions of $M$-theory (in a certain kinematic limit) have been proposed \cite{mtheory, taylor}. Fuzzy spaces emerge naturally as classical solutions of such models. Matrix models also appear, because they contain brane-like configurations, in elaborations of the gauge-gravity duality \cite{mald}. Analyses of gauge theories dimensionally reduced to matrix models have been useful in probing this duality.
Noncommutative spaces also appear in string theories in certain backgrounds with a constant nonzero value for the two-form gauge field \cite{taylor, DN}.

Fuzzy spaces are also closely related to the quantum Hall effect \cite{KNR}. For the classic Landau problem of a charged particle in a magnetic field, the corresponding energy spectrum consists of equally spaced Landau levels; each Landau level is degenerate and the energy gap separating consecutive levels is  proportional to the magnetic field $B$. For strong magnetic fields, the low energy physics is confined to the states within one, say the lowest Landau level (LLL).
The observables relevant for low energies are 
hermitian operators on this subspace of the Hilbert space; they are
given by the projection of the full operators to the lowest Landau level. The
operators representing coordinates, for example, when projected to
the LLL (or any other level), are no longer
mutually commuting. The LLL thus becomes a model of the noncommutative two-plane. (The appearance of 
noncommutativity in the string context mentioned above is similar, with the two-form field playing the role of the magnetic field.) Generalizing beyond the plane, for the Landau problem on a compact space of finite
volume, we get a finite number of states in the LLL, and the resulting 
subspace can be identified as ${\cal H}_N$, one of the ingredients for a fuzzy space.
Observables then become $(N\times N)$-matrices and there are natural choices for the Laplacian. More specifically, the LLL states for quantum Hall effect
on a space ${\cal M}$ gives us a fuzzy version of ${\cal M}$.

The main advantage of this point of view is that
the quantum Hall system gives us a model and a physical context to think about many issues related to fuzzy spaces. The lowest Landau level gives us a realization of the fuzzy space;
subspaces, specified by a projection operator, will correspond to Hall droplets.
The edge excitations of the Hall droplet describe the dynamics of the embedding of a disc
into the fuzzy space.
The bulk dynamics of the Hall droplet is related to the dynamics of gauge fields corresponding to  isometries of the fuzzy space, and hence, gravity.

In what follows, we will discuss such issues from both the matrix model-fuzzy space 
and the quantum Hall points of view. Thus all results can have two  different interpretations.
We will start with the quantum Hall effect since this gives a familiar physical
context.

\section{Quantum Hall effect in higher dimensions}

Quantum Hall effect in two dimensions is a very special physical phenomenon
which has led to an enormous amount of theoretical and experimental research \cite{2dqhe}. The basic phenomenon refers to the dynamics of charged fermions (electrons in a solid)
in a plane with a constant magnetic field orthogonal to it. At the single particle level, the energy eigenstates are grouped into the Landau levels. For high values of the magnetic field at low temperatures, the separation of levels is high compared to the available thermal excitation energy and the dynamics is confined
to the lowest Landau level. In a physical sample, there is also a potential $V$ which confines the fermions to within the sample. If we have $K$ fermions, they are localized near the minimum of $V$, but spread out over an area proportional to $K$ due to the exclusion principle. We get an incompressible droplet. Physically interesting issues are the bulk dynamics of the droplet, which refers to its response to changes in the externally applied
electromagnetic fields, and the edge dynamics which describes the fluctuations of the edge of the droplet. The electric current in the planar direction orthogonal to an applied
in-the-plane electric field, the so-called Hall current, is quantized, hinting at topological robustness in the underlying dynamics. As a result, there are many interesting mathematical facets to the theory.

The quantum Hall effect was generalized to the four-dimensional sphere
$S^4$ by Zhang and Hu \cite{ZH}. Since then further generalizations and analyses in higher dimensions and different geometries have been carried out by many authors \cite{KN1}-\cite{poly3}. The general framework is the following.
For any coset manifold of the $G/H$ type, where
$G$ is a Lie group and $H$ a compact subgroup
(of dimension $\geq 1$), the spin connection gives the analog
of a constant background field.
Thus it is possible to consider QHE on such spaces
taking the gauge field to be proportional to the
spin connection. In two dimensions,
one can consider $S^2 = SU(2)/U(1)$ which admits a constant
$U(1)$ background field and leads to the usual QHE on a two-sphere.
For $S^4 = SO(5)/SO(4)$, the isotropy
group is $H= SO(4)\sim SO(3) \times SO(3)$ giving the
possibility
of selfdual and antiselfdual fields, the instantons. This was the case 
considered by Zhang and Hu \cite{ZH}.
For ${\bf CP}^k = SU(k+1)/U(k)$, one can get constant background fields which
are either
abelian ($U(1)$) or nonabelian ($U(k)$). 
Other interesting cases which have been studied include
$S^3 = SU(2)\times SU(2) /SU(2)$ \cite{NR1}, the eight-sphere $S^8$ \cite{S8} and hyperbolic spaces based on noncompact groups \cite{jellal}.

The quintessential example for us is ${\bf CP}^k$, since it has all the characteristics we need and most of the other spaces which have been studied are special cases of this.
The case of $S^4$ can be recovered  from  QHE on
${\bf CP}^3$ since ${\bf CP}^3$ is an $S^2$-bundle
over $S^4$. As a result, ${\bf CP}^3$ with a $U(1)$ field leads to $S^4$ with a self-dual $SU(2)$ field as the background gauge field \cite{KN2}.
Likewise, since ${\bf CP}^7$ is a ${\bf CP}^3$-bundle over $S^8$, QHE on $S^8$ can be obtained from ${\bf CP}^7$ \cite{S8}.
The case of $S^3$ can be related to ${\bf CP}^1 \times {\bf CP}^1 = S^2\times S^2$
via the angle-axis embedding of $S^3/{\mathbb Z}_2$ in $S^2 \times S^2$ \cite{NR1}.
So, in short, we can use ${\bf CP}^k$ to formulate our calculations.
Most of the results, of course, will be generic.

\section{Quantum Hall effect on  ${\bf CP}^k$}

In this section, we shall consider the states in the lowest Landau level for the space 
${\bf CP}^k = SU(k+1) / U(k)$ \cite{KN1, KN2}; this space will be adequate for our considerations.

The symmetries of ${\bf CP}^k$ form the group $SU(k+1)$, with $U(k)$ as the local isotropy
group. The Riemannian curvature of ${\bf CP}^k$ takes values in the Lie algebra of 
$U(k)$, and because this is a homogeneous space, the curvature is constant in the basis of the frame fields. We can thus choose values of the background gauge field to be proportional to the curvature; this would give us a generalization of the ``constant magnetic field''.
The Landau problem is defined by this choice of magnetic field and one can then solve for the Landau levels.

The construction of the wave functions for the Landau levels can be done as follows.
Let $g$ denote a general element of $SU(k+1)$ in the fundamental representation;
i.e., it is a $(k+1)\times (k+1)$ matrix. The representative of $g$ in a representation $J$ is the Wigner ${\cal D}$-function
corresponding to that representation. If ${\hat g}$ denotes a general operator version of
$g$, then we may write the $\D$-function as
\beq
\D^{(J)}_{L,R} (g) ~=~ \la J, l\vert ~{\hat g}~\vert J, r\ra
\label{diff1}
\eeq
where $l,~r$ label the states within the representation $J$.
Functions on $SU(k+1)$ can be expanded in a basis of the $\D$-functions; functions on
${\bf CP}^k = SU(k+1)/U(k)$ are given by functions on $SU(k+1)$ which are $U(k)$-invariant.

We define the  left and right translation operators on $g$ by
\beq
L_A ~g = t_A ~g, \hskip 1in R_A ~g = g~t_A
\label{diff2}
\eeq
Here, $t_A$,
$A = 1, 2, \cdots, k^2+2k$, are a set of hermitian matrices which form a basis of the Lie algebra of $SU(k+1)$ in the fundamental representation. These are taken to obey
\beq
[ t_A, t_B ] = i f_{ABC} t_C , \hskip 1in \Tr (t_A t_B )= \half \delta_{AB}
\label{diff3}
\eeq
$f_{ABC}$ are the structure constants of $SU(k+1)$ in this basis.
The right translation operators can be split into the subgroup and coset generators as follows.
$R_{k^2+2k}$ will denote the $U(1)$ generator
in $U(k)\subset SU(k+1)$, $R_a$, $a= 1, 2, \cdots, k^2-1$, will denote $SU(k)$ generators.
The coset components which are in the complement of
$\underline {U(k)}$ in the Lie algebra $\underline{SU(k+1)}$ will be denoted by
$R_{\alpha}$, $\alpha=1, 2, \cdots, 2k$. The coset generators can be further separated into the raising and lowering type $R_{\pm I}= R_{2I-1} \pm i R_{2I}$, $I =1, \cdots, k$. (A similar splitting can be made for the left translations, but they will not be needed for what follows.)

The translation operators $R_A,~L_A$ can be realized as differential operators with respect to the parameters of $g$. The coset operators $R_{\alpha}$ correspond to covariant derivatives  while the $SU(k+1)$ operators $L_A$ correspond to magnetic translations. In particular
the covariant derivatives on ${\bf CP}^k$ can be taken to be
$D_{\pm I}= i R_{\pm I} /R$, where $R$ is a scale factor giving the radius of ${\bf CP}^k$.
Since $[R_{\pm I}, R_{\pm J}] \in \underline {U(k)}$, we get $[R_{\pm I} , R_{\pm J}]~f =0$
for functions $f$ on ${\bf CP}^k$ since they are $U(k)$-invariant.
The commutator of the covariant derivatives on the wave functions of charged particles must be proportional to the field strength. Thus they will not be true functions in ${\bf CP}^k$ but rather sections of a bundle. We consider a general background where there is 
a constant $U(1)$ field proportional to the $U(1)$-component of the curvature 
and a constant nonabelian $SU(k)$ field proportional to the $SU(k)$ component of the curvature. The particles will be taken to have a unit abelian charge and to transform as a representation $J'$ of $SU(k)$ for the nonabelian part.
The statement about background fields can then be
encoded in the commutation rules if we require the wave functions to obey
\beqar
R_a ~\Psi_{m; a'}  &=& 
\Psi_{m;b'} ~ (T_a)_{b' a'} \nonumber \\ 
{R}_{k^2 +2k} ~\Psi_{m; a'} &=& - {n k \over \sqrt{2 k
(k+1)}}~\Psi_{m; a'} 
\label{diff4}
\eeqar
The indices $a' ,b'~=1,\cdots, N'$ label the states within the $SU(k)$ representation $J'$. 
The matrices $T_a$ are the $SU(k)$ generators in the representation $J'$.
For a unitary realization of the right translations, $T_a$ should be the generators of
a unitary representation of $SU(k) \in SU(k+1)$ and, for the $U(1)$ part, $n$ has to be an integer, so that
the $U(1)$ action is part of a unitary representation of $SU(k+1)$. These are Dirac-type quantization conditions.
In the special case when there is no $SU(k)$ field, these simplify as
\beqar
 R_a ~\Psi_{m} &=& 0\nonumber \\
{R}_{k^2 +2k} ~\Psi_{m} &=& - {n k\over \sqrt{2 k
(k+1)}}~\Psi_{m} 
\label{diff5}
\eeqar
The wave functions obeying these conditions will be proportional to $\D^{(J)}_{L,R}$, where
the state $\vert J, r\ra$ is chosen to have the eigenvalue $- {n k/ \sqrt{2 k
(k+1)}}$ for the $U(1)$ generator $T_{k^2+2k}$ and to transform as the $J'$ representation of
$SU(k)\in SU(k+1)$. The representation $J$ of $SU(k+1)$ must be so chosen that
it contains such an $SU(k)$
representation, with the assigned $U(1)$ charge.

The Laplacian for the space is given by $-\nabla^2 = R_{+I}R_{-I} + R_{-I} R_{+I}
= 2 R_{+I}R_{-I} + {\rm constant}$. The Hamiltonian for the Landau problem
will be proportional to this for the nonrelativistic case and proportional to
$\sqrt{-\nabla^2 +m^2}$ for the relativistic case; in any case, it is an increasing function of
$R_{+I}R_{-I}$. We see that the minimum of the Hamiltonian, and hence the lowest Landau level, is
given by wave functions obeying 
\beq
R_{-I} ~\Psi_{m;a'} =0
\label{diff6}
\eeq
This means that, for the lowest Landau level, in addition to the conditions (\ref{diff4}), 
$\vert J, r\ra$ must be a lowest weight state with $T_{-I} \vert J, r\ra =0$;
we will denote these states as $\vert a', -n\ra$.
Once the representation $J'$ is specified, one can identify representations $J$ of
$SU(k+1)$ which contain such a state. For example, if
there is no $SU(k)$ field, the symmetric rank $n$ representation of $SU(k+1)$
will contain the lowest weight state $\vert -n\ra$, which is an $SU(k)$ singlet.
The properly normalized wave functions are given by
\beq
\Psi_{m; a'} (g) = \sqrt{N} ~\la J, l \vert ~{\hat g} ~\vert a', - n\ra
~\equiv \sqrt{N}~ {\cal D}_{m;a'}(g)
\label{diff7}
\eeq
where $N$ is the dimension of the representation $J$ of $SU(k+1)$. These are normalized
by virtue of the orthogonality theorem
\beq
\int d\mu (g) ~\D^*_{m;a'} (g)~\D_{k;b '}
(g) ~=~ {\delta_{mk}\delta_{a' b '}\over N}
\label{diff8}
\eeq

It is instructive to relate this group-theoretic analysis to the standard 
discussion of ${\bf CP}^k$ in terms of homogeneous and local coordinates.
We begin by recalling that
${\bf CP}^k$ is a $2k$-dimensional manifold parametrized by $k+1$ complex coordinates $v_a$, such that $\bar{v}_a v_a =1$, 
with the identification $v_a \sim e^{i \theta} v_a$. One can further introduce local complex coordinates $z_{I}$, $I=1,\cdots, k$, by writing
\beq
v_{I}  =  {z_{I} \over {\sqrt{1 + \bz \cdot z}}}~,\hskip .3in 
v_{k+1}  =  {1 \over {\sqrt{1 + \bz \cdot z}}} 
\label{complex}
\eeq

We can now use a group element $g$ in the fundamental representation of $SU(k+1)$ to parametrize ${\bf CP}^k$, by making the identification $g \sim gh$, where $h \in U(k)$. 
We can use the freedom of $h$ transformations to write $g$ as a function of the real coset coordinates $x^i$, $i=1,\cdots,2k$. The relation between the complex coordinates $z^I,~\bar{z}^I$ in (\ref{complex}) and $x^i$ is the usual one, $z^I = x^{2I-1} + i x^{2I},~I=1,\cdots,k$. 
The homogeneous coordinates are related to the group element by
$g_{I,k+1} = v_I$, $g_{k+1,k+1} = v_{k+1}$.

For the variation of $g$, we can write
\beq
g^{-1}dg = \big( -i E^{k^2+2k}_i t_{k^2+2k} -i E^{a}_i t_{a} -i E^{\alpha}_i t_{\alpha} \big)~ dx^i
\label{gdg}
\eeq
The $E^{\alpha}_i$ are the frame fields in terms of which the Cartan-Killing metric on ${\bf CP}^k$ is given by
\beq
ds^2 = g_{ij} dx^i dx^j = E^\alpha_i E^\alpha_j dx^i dx^j
\label{metric}
\eeq
The K\"ahler two-form on ${\bf CP}^k$ is likewise written as
\beqar
\omega_K & = & -i \sqrt{{2k \over {k+1}}} \tr \left( t_{k^2 + 2k} ~ g^{-1}dg~  g^{-1}dg \right)
\nonumber
\\ & = & -{1 \over 4} \sqrt{{2k} \over {k+1}} f^{(k^2+2k)\alpha\beta} ~E^\alpha_i~ E^{\beta}_j ~dx^i \wedge dx^j ~\equiv ~{1 \over 2} (\omega_K) _{ij}~ dx^i
\wedge dx^j
\label{18d}
\eeqar
The fields $E^{k^2+2k}_i$ and $E^a_i$ are related to the $U(1)$ and  $SU(k)$ background gauge fields on ${\bf CP}^k$. In particular the $U(1)$ field $a$ is given by
\beq
a  =  i n \sqrt{{{2k} \over {k+1}}} \tr (t_{k^2+2k} g^{-1} dg ) = {n \over 2} \sqrt{{{2k} \over {k+1}}} E^{k^2+2k} %\nonumber \\
%& = & -i n~g^*_{k+1,a}dg_{a, k+1}
\label {Abar}
\eeq
We can similarly define an $SU(k)$ background field $\bar{A}^a_i$. Its normalization is chosen so that
\beq
\bar{A}^a \equiv E^a =  2i \tr (t^a g^{-1} dg) \label{NA}
\eeq
Notice that $\bar{A}^a$ in (\ref{NA}) does not depend on $n$, while the abelian field $a$ in (\ref{Abar}) is proportional to $n$. 
The corresponding $U(1)$ and $SU(k)$ background field strengths are
\beqar
\del_i a_j - \del_j a_i  &=&  n (\omega_K)_{ij} = - {n \over 2} \sqrt{{{2k} \over {k+1}}} f^{(k^2+2k)\alpha\beta} E^{\alpha}_iE^{\beta}_j  \nonumber \\
\bar{F}^{a}_{ij}  &=&  \del_i \bar{A}^a_j -\del_j \bar{A}^a_i + f^{abc} \bar{A}^b_i \bar{A}^c_j = - f^{a\alpha\beta} E^{\alpha}_iE^{\beta}_j
\label{barF}
\eeqar
We see from (\ref{barF}) that in the appropriate frame basis the background field strengths are constant, proportional to the $U(k)$ structure constants. It is in this sense that the field strengths in (\ref{barF}) correspond to uniform magnetic fields appropriate in defining QHE. 
The Maurer-Cartan equations 
\beq
d E^\alpha~-~ ( f^{a\alpha\beta} E^a + f^{k^2+2k ~\alpha\beta}E^{k^2+2k} )
~E^\beta =0
\label{diff9}
\eeq
show that the spin connections are given by
$- f^{a\alpha\beta} E^a$ and $-f^{k^2+2k ~\alpha\beta}E^{k^2+2k}$; the field strengths
(\ref{barF}) are thus proportional to the Riemann curvature of ${\bf CP}^k$.

The $U(1)$ background magnetic field (which leads to the Landau states) can be written in terms of the homogeneous coordinates as $a = -i n  \bar{v} \cdot dv$
with the field strength
\beq
 d a = -in d\bar{v} \cdot dv = n \omega_K
\label{18}
\eeq
We can also write
$n=2 B R^2$, in terms of the radius $R$ of
${\bf CP}^k$, identifying $B$ as the local value of the constant $U(1)$ magnetic field.
(The case of charged fermions on ${\bf CP}^1 =S^2$ with $U(1)$ background field,
corresponding to $k=1$, was studied by Haldane several years ago \cite{haldane}. In this case the background gauge  field $a$ is that of a monopole of charge $n$ placed at the origin of $S^2$.)

If there is only the $U(1)$ field, the representations
of $SU(k+1)$ which are relevant  for the lowest Landau level are totally
symmetric and are of rank $n$. 
The wave functions are then explicitly given in local coordinates as
\beqar
\Psi_m(\vec{x}) &=& \sqrt{N} \left[ {n! \over i_1! i_2! ...i_k! (n-s)!}\right]^{1\over 2} ~ {z_1^{i_1} z_2^{i_2}\cdots z_k^{i_k}\over
(1+\bz \cdot z )^{n / 2}}~,\qquad m=1,\cdots, N  \nonumber\\
~\\
s &=& i_1 +i_2 + \cdots +i_k ,\qquad 0\le i_i \le n~, \qquad  0 \le s \le n \label{wav}
\eeqar
These are the coherent states for ${\bf CP}^k$.
The number of states in the lowest Landau level is given by the
dimension of the symmetric rank $n$ representation as
\beq
N={\rm dim} J = {{(n+k)!} \over {n! k!}} 
\label{dim}
\eeq
Notice that, for large $n$, this gives $N\rightarrow n^k /k!$.
When there is a nonabelian background as well, the
dimension $N$ of the $SU(k+1)$ representation $J$ depends on the particular $J'$ representation chosen.  While the full formula depends on the details of
the representations, for large $n$, we have
\beq
N = {\rm dim} J \rightarrow {\rm dim} J' ~{n^k \over k!} = N' ~{n^k \over k!}
\label{35a}
\eeq

\section{Matrix formulation of quantum Hall (phase space) dynamics}

We are now in a position to present a matrix formulation of the dynamics of noninteracting fermions in the lowest Landau level, with and without external gauge interactions.  Our analysis in this section will be quite general, not necessarily restricted to ${\bf CP}^k$.

We consider $K$ fermions which occupy $K$ states out of the $N$ available states in the LLL.
The confining potential ${\hat V}$ lifts the degeneracy of the LLL states and the fermions are localized around the minimum of  ${\hat V}$ forming a droplet. Because of the exclusion principle and the conservation of the number of fermions, the excitations are deformations of the droplet which preserve the total volume of occupied states (volume of phase space). 

The droplet is mathematically characterized by a diagonal density matrix $\hat {\rho}_0$ which is equal to 1 for occupied states and zero for unoccupied states. Further,
$\hat{\rho}_0$ may be taken to be the density matrix for the many-body ground state. The most general fluctuations which preserve the LLL condition and the number of occupied states are unitary transformations of $\hat {\rho}_0$, namely $\hat {\rho}_0 \rightarrow \hat{\rho}=\hat{U}  \hat {\rho}_0 \hat{U} ^ \dagger$, where  $\hat{U}$ is an $(N \times N)$ unitary matrix representing the dynamical modes. One can write an action for these modes as
\beq
S_{0}=  \int dt~ \Tr \left[ i  {\hat \rho}_0 { \hat U}^\dagger \del_t {\hat U}
~-~ {\hat \rho}_0 {\hat U}^\dagger {\hat{V}} {\hat U} \right]
\label{1}
\eeq
where $\hat {V}$ is the confining potential. (The Hamiltonian is $\hat{V}$ up to an additive constant.) The unitary matrix $\hat{U}$ can be thought of as a collective variable describing all the possible excitations within the LLL. The equation of motion resulting from (\ref{1}) is the expected quantum Liouville equation for the density matrix $\hat{\rho}$,
\beq
i {{\partial \hat{\rho}} \over {\partial t}} = \left[ \hat{V},~ \hat{\rho} \right]
\label{3}
\eeq
The action $S_{0}$ can also be written as \cite{KN2}
\beq
S_{0}= {N \over N'}  \int d\mu dt~ \tr~\left[ i ({\rho}_0 *{  U}^\dagger  * \del_t { U})
~-~ ({ \rho}_0 *{U}^\dagger  * {{V}} * {U}) \right]
\label{4a}
\eeq
where $d\mu$ is the volume measure of the space where QHE has been defined and $\rho_0,~U,~V$ are the symbols of the corresponding matrices on this space. (The hatted expressions correspond to matrices and unhatted ones to the corresponding symbols, which are fields on the space where QHE is defined.)  Equation (\ref{4a})
is written for the case of nonabelian fermions coupled to a background gauge field in some representation $J'$ of dimension $N'$; the corresponding symbols are $(N' \times N')$ matrix valued functions.
We will use ``$\Tr$" to indicate the trace over the $N$-dimensional LLL Hilbert space while ``$\tr$" indicates trace over the $N'$-dimensional representation $J'$. In the case of abelian fermions, $N'=1$ and $\tr$ is absent. (The large $N$ limit we are considering will keep
$N'$ finite as $N\rightarrow \infty$.)

The general definitions of the symbol and the star-product are as follows.
If $\Psi_m(\vec{x})$, $m=1,\cdots,N$,  represent the correctly normalized LLL wave functions, then the symbol corresponding to an $(N \times N)$-matrix $\hat{O}$, with matrix elements $O_{ml}$ is
\beq
O(\vec{x}, t) = { 1 \over N} \sum_{m,l} \Psi_m(\vec{x}) ~O_{ml}(t) ~\Psi^*_l(\vec{x})
\label{symb}
\eeq
The star-product is defined by the condition that the symbol for the product of two matrices is given as the star-product of the symbols for the individual matrices, i.e., by
$\big(\hat{O}_1\hat{O}_2\big)_{symbol} = O_1(\vec{x}, t) * O_2 (\vec{x}, t)$.

Notice that the dynamics of the underlying fermion problem is described in terms of a one-dimensional matrix action (\ref{1}), which can also be written as a noncommutative field theory action, as in (\ref{4a}). The matrices and the action in (\ref{1}) do not depend on the particular space and its dimensionality or the abelian or nonabelian nature of the underlying fermionic system. This information is encoded in equation (\ref{4a}) in the definition of the symbol, the star-product and the measure.

This matrix formulation can be extended to include external gauge fields
which are in addition to the uniform background magnetic field which defines the Landau problem \cite{KA2}.
These additional fields will be often referred to as the gauge field fluctuations.
Gauge interactions should be described by a matrix action $S$ which is invariant under time dependent $U(N)$ rotations, $ \hat{U} \rightarrow \hat{h} \hat{U}$,
where $\hat{h} = \exp (-i \hat{\lambda})$ for some hermitian matrix ${\hat \lambda}$.
The action will be the gauged version of $S_0$, with $\del_t$ replaced by 
the covariant derivative $\hat{D}_t= \del_t + i \hat{\A}$, where $\hat{\A}$ is a matrix gauge potential. Thus,
\beq
S= \int dt~ \Tr \left[ i  \hat {\rho}_0  \hat {U}^\dagger ( \del_t + i \hat{\A} ) \hat {U}
~-~  \hat {\rho}_0 \hat {U}^\dagger \hat{V} \hat{U} \right]
\label{6}
\eeq
Invariance of this action under infinitesimal time dependent $U(N)$ rotations 
$\delta\hat{U} = - i \hat{\l} \hat{U}$
implies the following transformation for the gauge potential $\hat{\A}$,
\beq
\delta \hat{\A}  =   \del _t {\hat{ \l }}-i [\hat{\l}, \hat{V}+\hat{ \A} ]
\label{8}
\eeq
The action (\ref{6}) can be written in terms of the corresponding symbols as
 \beq
S= {N \over N'}   \int dt~d\mu ~ \tr~\left[\rho_0 *\left( i   U^\dagger  * \del_t U
~- U^\dagger * V *U- U^\dagger * \A *U \right)\right]
\label{10}
\eeq
The action (\ref{10}) is now invariant under the infinitesimal
transformations
\beqar
\delta U  &=& - i \l * U \nonumber \\
\delta \A   &=&  \del _t \l  - i \left( \l* (V+\A) - (V + \A) * \l \right) 
\label{11}
\eeqar
We shall refer to this as the $W_N$-gauge transformation, in analogy to the $W_\infty$ transformation appearing in the case of the planar two-dimensional QHE \cite{IKS, wadia}.

The key physical question is how the field $\A$ is related to the gauge fields $A_\mu$ to which the fermions couple in the usual way.
Once this is known, the 
action (\ref{10}) can be expressed in terms of the usual gauge fields.
For the gauge interactions of the original fermion system, we have invariance
under the usual gauge transformation
 \beq
 \delta A_{\mu}  =  \del_{\mu} \Lambda + i [\bar{A}_\mu + A_\mu , ~ \Lambda] ,~~~~~~~
 \delta \bar{A}_{\mu}  =  0 \label{13}
 \eeq
Here $\Lambda$ is the infinitesimal gauge parameter and $\bar{A}_\mu$ is the nonabelian uniform background field. What we need is an expression for $\A$ in terms of
$A_\mu$ such that when the gauge fields $A_\mu$ are transformed as 
in (\ref{13}), the field $\A$ undergoes the transformation (\ref{11}).
In other words, the transformation (\ref{11}) is induced by the transformation
(\ref{13}). This is the basic principle
which can be used to
determine $\A$ as a function of $A_{\mu}$, up to gauge invariant terms.
The bosonized action of the LLL fermionic system in the presence of gauge interactions
then follows in a straightforward way.
 Since ${\A}$ is the time component of a noncommutative gauge field, the relation 
 between ${\A}$ and the commutative gauge fields $A_\mu$ is essentially a Seiberg-Witten transformation \cite{seib1,seib2}.
 
It is quite clear that the possible excitations of the LLL fermionic system are particle-hole excitations, which can, in principle, be described in terms of bosonic degrees of freedom. The noncommutative field theories given by the actions  
 $S_0$ in (\ref{4a}) and $S$ in (\ref{10}) are the exact bosonic actions describing the dynamics of the noninteracting LLL fermions without or with gauge interactions. 
 Thus the matrix theory provides a very general way to construct the 
 bosonic action for a fermionic system by viewing it in phase space as a Landau problem
 with the symplectic structure being the magnetic field. Some of these ideas have already been used in the context of phase space bosonization for one-dimensional nonrelativistic fermions \cite{wadia} and for the effective droplet dynamics in the planar quantum Hall effect \cite{sakita, shizuya}.

We shall now demonstrate that in the limit where $N \rightarrow \infty$ and the number of fermions is large, the action $S_0$ reduces, for arbitrary even dimensions, to a boundary action describing the edge excitations (abelian and nonabelian) of the QHE droplet \cite{KN2}. In the presence of fluctuating gauge fields there is an additional bulk action, given in terms of a Chern-Simons term, whose anomaly gets cancelled by the boundary contribution given in terms of a generalized chiral, gauged Wess-Zumino-Witten action \cite{KA2}. 

\section{Star product for ${\bf CP}^k$ with $U(k)$ background gauge field}

The large $N$ simplifications are carried out using the symbols and star-products.
Let $\hat{X}$ be a general $(N \times N)$-matrix, with matrix elements $X_{ml}$, acting on the $N$-dimensional Hilbert space generated by the basis (\ref{diff7}). The symbol corresponding to $\hat{X}$ is defined by
\beqar
X_{a'b'} (\vec{x}, t) &  =  & {1 \over N} \sum_{ml}{\Psi}_{m;a'}(\vec{x}) ~X_{ml} (t)  {\Psi}^*_{l;b'}(\vec{x}) \nonumber \\
& = & \sum_{ml}{\cal D}_{m;a'}(g) ~X_{ml} {\cal
D}^*_{l;b'}(g) 
~ =  \la b',-n \vert {g}^{\dagger} X^T {g} \vert a';-n \ra
\label{symbolNA}
\eeqar
In the nonabelian case the symbol is a $(N' \times N')$ matrix valued function, while in the abelian case where $J'$ is the singlet representation, the symbol is just a function on ${\bf CP}^k$. With this definition
\beq
\Tr {\hat X}  =  {N \over N'} \sum_{a'} \int d\mu (g) ~X_{a'a'} (g)
\label{}
\eeq
The symbol corresponding to the product of two matrices $\hat{X}$ and $\hat{Y}$ is given by the star-product of the symbols for $\hat{X}$ and $\hat{Y}$, i.e., 
\beqar
(\hat{X}\hat{Y})_{a'b'} & = & X_{a'c'} *  Y_{c'b'}                                
~ =  \sum_{mrl} {\cal D}_{m;a'}(g) ~X_{mr}Y_{rl}
{\cal D}^{*}_{l;b'}(g) \nonumber \\
& = & \la b',-n \vert {g}^\dagger Y^T X^T {g} \vert a',-n \ra 
~  =  \la b',-n \vert {g}^\dagger Y^T ~ {\bf 1}~X^T {g} \vert a',-n \ra
%\nonumber\\
%&\equiv& \sum_{\gamma} A_{\alpha \gamma} * B_{\gamma \beta}
\label{star}
\eeqar
In order to evaluate the star-product we need to reexpress the unit matrix ${\bf 1}$ in (\ref{star}), where ${\bf 1} = \sum_m \vert m \ra \la m \vert$, and $\vert m \ra$ are all the states in the $J$ representation, in terms of the lowest weight states $\vert a',-n \ra$.  In the case 
of a $U(1)$ background field the star-product, following this method, was derived in \cite{KN2}. We found
\beq
X*Y = \sum_s (-1)^s \left[ {(n-s)! \over n! s!}\right]
\sum_{\sum i_k=s}^n
{s! \over i_1! i_2! \cdots i_k!}~\bigl({ R}_{-1}^{i_1} 
\cdots { R}_{-k}^{i_k} X \bigr)
~\bigl({ R}_{+1}^{i_1} \cdots { R}_{+k}^{i_k}Y\bigr)
\label{25a}
\eeq
Expression (\ref{25a}) can be thought of as a series expansion in $1/n$. Similar expressions for the star-product of functions were derived in the context of noncommutative ${\bf CP}^k$ \cite{bal5}.

In the case of the $U(k)$ background field the calculation of the star-product to arbitrary order in $1/n$ is very involved. The calculation to order $1/n$ was done in
\cite{KN2} and extended to order $1/n^2$ 
in \cite{KA2}. The result is
\beqar
X*Y &=& XY - { 1 \over n} R_{-I}X R_{+I}Y + {i \over n^2} R_{-J} X f^{a \bar{I}J} (T_a)^{T} R_{+I} Y \nonumber \\
&&+  {1 \over {2n^2}} R_{-I}R_{-J} X R_{+I} R_{+J} Y~+~ {\Or}({1/ n^3})
\label{43}
\eeqar

The right translation operators $R_{\alpha}$ can be expressed as differential 
operators using (\ref{diff2}) and 
(\ref{gdg}),
\beqar
R_\alpha {g}  &=&  i (E^{-1})^i_\alpha \big( \del_i {g} + i {g} E_i^{k^2+2k} T_{k^2+2k} +i {g}
E_i^a T_a \big)~ \equiv i (E^{-1})^i_\alpha D_i {g} \nonumber\\
R_\alpha {g}^\dagger  &=&  i (E^{-1})^i_\alpha \big( \del_i {g}^\dagger - i  E_i^{k^2+2k} T_{k^2+2k} {g}^\dagger -i  E_i^a T_a {g}^\dagger \big)~ \equiv i (E^{-1})^i_\alpha D_i {g}^\dagger 
\label{44}
\eeqar
where $T$'s are the $U(k)$ generators in the particular representation ${g}$ belongs to. Using (\ref{44}) and the symbol definition (\ref{symbolNA}), we find that the action of $R_\alpha$ on a symbol is
\beqar
R_\alpha X_{a'b'} & = & i (E^{-1})^i_\alpha ( D_i X)_{a'b'} \nonumber \\
D_i X & = & \del_i X + i [\bar{A}_i ,~X],~~~~~~~~~~~\bar{A}_i = \bar{A}^a_i (T_a)^T= E_i^a (T_a)^T
\label{44a}
\eeqar
where $\bar{A}$ is the $SU(k)$ background gauge field in the $J'$ representation. Notice that the $U(1)$ part of the gauge field does not contribute in (\ref{44a}). 

Combining expressions (\ref{43}) and (\ref{44a}) we can rewrite the star-product in terms of covariant derivatives and real coordinates (instead of complex) as
\beqar
X*Y  &=&  XY + {1 \over n} P^{ij} D_i X D_j Y - {i \over n^2} P^{il} P^{kj} D_i X \bar{F}_{lk} D_j Y \nonumber \\
&&+  { 1 \over {2 n^2}} P^{ik} P^{jl} {\cal{D}}_iD_j X {\cal{D}}_kD_l Y + {\Or}({1 / n^3})
\label{47}
\eeqar
where $\bar{F}_{lk} = \bar{F}^a_{lk} (T_a)^T$ and
$P^{ij} = g^{ij} + {i \over 2} (\omega_K^{-1})^{ij}$. ${\cal{D}}_i$ is the Levi-Civita covariant derivative for a curved space 
such as ${\bf CP}^k$, namely,
\beqar
{\cal{D}}_iD_j X & \equiv & D_iD_j X - \Gamma^l_{ij} D_l X \nonumber \\
D_i E^\alpha_j & = & \del_i E_j^{\alpha} + f^{\alpha A \beta} E_i^A E_j^\beta ~ = ~ \Gamma_{ij}^l E_l^\alpha
\label{46}
\eeqar
where $A$ in $f^{\alpha A \beta}$ is a $U(k)$ index (both $U(1)$ and $SU(k)$) and $\Gamma_{ij}^l$ is the Christoffel symbol for ${\bf CP}^k$. 

Equation (\ref{47}) is valid for both abelian and nonabelian cases. In the abelian case, they simplify since $ X,~Y $ are commuting functions and $\bar{F}_{lk} \rightarrow 0$, so that $D_i X \rightarrow \del_iX$ and ${\cal{D}}_iD_j X \rightarrow  \del_i\del_j X -\Gamma_{ij}^l \del_l X$.

\section{Calculation of ${\A}$}

In this section, we will outline the calculation of $\A$ as a function of $A_\mu$ via the
implementation of 
the $W_N$ transformation (\ref{11}) as induced by the gauge transformation (\ref{13})
on $A_\mu$.
Using (\ref{11}) and (\ref{47}) we find that up to $1/n^2$ terms
\beqar
\delta \A  &=&  \del_t \lambda - i[\lambda,~V+\A] - { i \over n} P^{ij} \left( D_i \lambda D_j (V + \A) - D_i (V + \A) D_j \lambda \right)  \nonumber \\
&&- {1 \over n^2} P^{il} P^{kj} \left( D_i \lambda \bar{F}_{lk} D_j V - D_i V \bar{F}_{lk} D_j \lambda\right) \nonumber \\
&&- { i \over {2 n^2}} P^{ik} P^{jl} \left( {\cal{D}}_iD_j \lambda {\cal{D}}_kD_l V - {\cal{D}}_iD_j V {\cal{D}}_kD_l \lambda\right) \label{51}
\eeqar

At this stage, it is useful to
discuss the scaling of various quantities. All expressions so far (including the measure $d\mu$, $g^{ij}$, $(\omega_K^{-1})^{ij}$, etc.) have been written in terms of the dimensionless coordinates $x_i = \tilde{x}_i/R$, where $R$ is the radius of ${\bf CP}^k$ and $\tilde{x}$ are the dimensionful coordinates. The calculation of the star-product (\ref{47}) involves a series expansion in terms of $1/n$, where $n=2 B R^2$ and $B$ is the magnitude of the
constant $U(1)$ magnetic field. Written in terms of the dimensionful parameters $\tilde{x}_i$, the expansion in  $1/n$ becomes an expansion in $1/B$. We shall
further assume that the energy scale of the gauge field fluctuation
$A_\mu$, and therefore of $\A$, is much smaller than $B$ to be consistent
with the restriction to LLL. 

The scale of the confining potential $V$ is set by the magnetic field $B$ ($\sim n$ in terms of dimensionless variables). A convenient choice for the confining matrix potential $\hat{V}$ is such that the ground state density  $\rho_0(\vec{x})$ corresponds to a spherical droplet. This is the case when all the $SU(k)$ multiplets of the $J$ representation 
upto a fixed hypercharge (the eigenvalue of $T_{k^2+2k}$) are completely filled, starting from
the lowest. A simple choice for such a potential is the one used in \cite{KN2}, 
\beq
\hat{V} = \sqrt{2k \over k+1}~\nu ~\left( { T}_{k^2+2k} + {nk \over
\sqrt{2k(k+1)}}
\right) \label{52}
\eeq
where $\nu$ is a constant. (The potential does not have to be exactly of this form;
any potential with the same qualitative features will do.)  The particular expression (\ref{52}) has the property that $\la s \vert \hat{V} \vert s \ra = \nu s$,
where $\vert s \ra$ denotes an $SU(k)$ multiplet of hypercharge $-nk +sk + s$, namely
$\sqrt{2k(k+1)}{T}_{k^2+2k} \vert s \ra = (-nk +sk + s) \vert s \ra$. The symbol for (\ref{52}) was calculated in \cite{KN2} to be
\beq
V_{a'b'} \equiv \la b',-n \vert {g}^\dagger {V}^T  {g} \vert a',-n\ra
= \nu n {{\bar{z} \cdot z} \over {1 + \bar{z} \cdot z}} \delta_{a' b'} ~+~S_{k^2+2k,a} (T_a)_{b' a'}\label{54}
\eeq
where $S_{k^2+2k,a} = 2 \tr ({g}^\dagger t_{k^2+2k} ~{g}~ t_{a})$.
The important point is that the first term in (\ref{54}) is diagonal and of order $n$ in terms of the dimensionless variables $z$, while the second non-diagonal term is of order $n^0$. In analyzing (\ref{51}) we can absorb the order $n^0$ term of the confining potential in the definition of $\A$ and treat separately the diagonal term of order $n$ as 
the potential $V$ to be used for large $n$ simplifications.
In this way, since $V_{a'b'} = \delta_{a'b'} V(r)$, where $r^2=\bar{z}\cdot z$, is proportional to the identity,
expression (\ref{51}) can be further simplified as
\beqar
\delta \A  &=&  \del_t \lambda - i[\lambda,~\A]  - { i \over n} P^{ij} \left(D_i \lambda D_j \A - D_i \A D_j \lambda \right) \nonumber \\
&&+ u^i D_i \lambda  - {i \over n} \left(P^{il} D_i \lambda \bar{F}_{lk} - P^{li} \bar{F}_{lk} D_i \lambda \right) u^k \nonumber \\
&&+ { 1 \over {2 n^2}} \left[(\omega_K^{-1})^{ik} g^{jl} + g^{ik}  (\omega_K^{-1})^{jl}\right]  {\cal{D}}_iD_j \lambda 
\nabla_k \del_l V
\label{56}
\eeqar
where
\beq
\nabla_k \del_l V = \del_k \del_l V - \Gamma^n_{kl} \del_n V, \qquad u^i = { 1 \over n} (\omega_K^{-1})^{ij} \del_j V
\label{56a}
\eeq
The quantity $u^i$ is essentially the phase space velocity, if we think of the LLL as the phase space of a lower dimensional system, with symplectic structure $n \omega_K$ and Hamiltonian $V$. 

Equation (\ref{56}) gives the transformation of $\A$. What we are seeking is
an expression for $\A$ as a function of $A_\mu$, $\A = f(A_\mu)$, such that 
\beqar
\delta \A ~({\rm as ~in~ eq.(\ref{56})}) &=&  f(\delta A_\mu) \label{57} \\
\delta A_\mu  &=&  \del_\mu \Lambda + i [\bar{A}_\mu + A_\mu, \Lambda]= D_\mu \Lambda + i [A_\mu, \Lambda]  , \qquad
\delta \bar{A}_\mu  =  0 \nonumber
\eeqar
The solution for $\A$ can be worked out from this
requirement, although the calculation is algebraically a bit tedious \cite{KA2}. It is given by
\beqar
\A &=&  A_0 -{i \over 2n} g^{ij} \left[ A_i,~2 D_i A_0 - \del_0 A_i + i [A_i,~A_0] \right]  \nonumber \\
&&+ { 1 \over {4n}} (\omega_K^{-1})^{ij} \{A_i, 2 D_j A_0 - \del_0 A_j + i [A_j,~A_0] \} \nonumber \\
&&+ u^i A_i -{i \over {2n}} g^{ij} \left[A_i,~A_k \right] \del_j u^k + {1 \over {4n}} (\omega_K^{-1})^{ij} \{A_i,~A_k \} \del_j u^k \nonumber \\
&&-{i \over {2n}} g^{ij} \left[ A_i,~  2 D_j A_k - D_k A_j + i [ A_j,~A_k] ~ + 2 \bar{F}_{jk} ~\right] u^k \nonumber \\
&&+ {1 \over {4n}} (\omega_K^{-1})^{ij} \{ A_i,~  2 D_j A_k - D_k A_j + i [ A_j,~A_k] + 2 \bar{F}_{jk} ~ \} u^k \nonumber \\
&&+ {1 \over {2n^2}} g^{ik} (\omega_K^{-1})^{jl} \left( {\cal{D}}_i A_j + {\cal{D}}_j A_i \right) \nabla_k\del_l V
\label{66}
\eeqar
where $[~,~]$ and $\{~,~\}$ indicate commutators  and anticommutators , respectively.
The symbol for the matrix gauge transformation parameter ${\hat\lambda}$ 
can also be evaluated as
\beq
\lambda = \Lambda - { i \over {2n}} g^{ij} \left[ A_i,~ D_j \Lambda \right] +{1 \over {4n}} 
(\omega_K^{-1})^{ij} \{A_i, 2 D_j \Lambda \}  + ~ {\Or} ( 1 / n^2 )  \label{59}
\eeq
The gauge field $A_\mu$ in (\ref{66}) contains both the abelian $U(1)$ and 
nonabelian $SU(k)$ components. 
In the abelian case where the fermions interact only with the $U(1)$ gauge field, the symbols are commuting functions, so the commutator terms in (\ref{66}) vanish. In terms of the dimensionful quantities $\tilde{x}= R x$ , $\tilde{D}= D/R,~\tilde{A}=A/R,~\tilde{V} \sim B$, $\A$ can be written as a series expansion in $1/B$.  The terms shown in
(\ref{66}) account for all terms of order $B^0$ and $1/B$.

The function $\A$ being the symbol of the time-component of the matrix gauge potential can be thought of as the Seiberg-Witten map \cite{seib1,seib2} for a curved manifold in the presence of nonabelian background gauge fields.

It is clear from (\ref{57}) that the expression (\ref{66}) is only determined up to gauge invariant terms whose coefficients are not constrained by the $W_N$-transformation (\ref{11}) and the requirement that it is induced via the gauge transformation (\ref{13}).  As we shall see in the next section, this solution produces the minimal gauge coupling for the chiral field describing the edge excitations of the quantum Hall droplet.

\section{Edge and bulk actions and anomaly cancellation}

The simplification of the action (\ref{10}) requires one more ingredient, namely, the symbol
for ${\hat \rho}_0$.
For the case of a confining potential $\hat{V}$ with an $SU(k)$ symmetry, as discussed in the previous section, one can perform an exact calculation for $\rho_0$. In the limit where $N$ is large and the number of fermions $K$ is large, where $N \gg K$, one can show that the symbol corresponding to the density matrix is of the form
\beq
(\rho_0)_{a'b'}   =  \rho_0(r^2) \delta_{a'b'} ,\quad
\rho_0(r^2) =  \Theta \Big(1 - {{R^2 r^2} \over {R_D^2}}\Big)
\label{step}
\eeq
where $\Theta$ is the step-function and $R_D$ is the radius of the droplet.  Equation (\ref{step}) defines the so-called droplet approximation for the fermionic density. We want to evaluate the action $S$, and identify the edge and bulk effective actions,
in this approximation. As mentioned earlier, the $1/n$ expansion of various quantities can be thought of as an expansion in $1/B$ if we write our expressions in terms of the dimensionful coordinates $\tilde{x}$. Similarly, using (\ref{35a}), the prefactor $(N/N') d\mu \rightarrow [n^k /(k! R^{2k})] d \tilde{\mu} = (2B)^k/k!~d\tilde{\mu}$, where $d\tilde{\mu}$ is the measure of the space in terms of the dimensionful coordinates. For convenience  we will continue the evaluation of  the edge and bulk effective actions in terms of the dimensionless coordinates, keeping in mind, though, that the $1/n$-expansion can always be converted to a $1/B$-expansion with the appropriate overall prefactor to  correctly accomodate the volume of the droplet.  

The large $n$ limit of the bosonic action $S_0$ in the absence of gauge interactions was derived in \cite{KN2}. It can be written in terms of a unitary 
$(N' \times N')$ matrix valued field $G= e^{i \Phi}$, where $\Phi$ is the symbol corresponding to $\hat{\Phi}$ in $\hat{U} = e^{i \hat{\Phi}}$. Explicitly
\beqar
S_0 &=& -{N \over {2nN'}} \int  dt d\mu ~{{\del \rho_0} \over \del r^2} ~
 \tr \left[ \left( G^{\dagger} {\dot G} + \nu ~G^{\dagger} D_{\omega} G \right)  
G^{\dagger}D_{\omega} G  \right] \nonumber \\
&&+ {{Nk} \over { 4 \pi nN'}} \int \rho_0 \left[ -d \left( i \bar{A} dG G^{\dagger} + i \bar{A} G^{\dagger}dG \right) + {1 \over 3} \left( G^{\dagger}dG \right)^3 \right] \wedge \left( {\omega_K \over {2 \pi}}\right)^{k-1} 
\label{70}
\eeqar
In this equation, $D_{\omega}$ is the component of the covariant derivative $D$ perpendicular to the radial direction, along a special tangential direction on the droplet boundary, given explicitly as
$D_{\omega} = - (\omega_K^{-1})^{ij} 2r \hat{x}_i D_j$; 
$\hat{x}_i$ is the radial unit vector normal to the boundary of the droplet.
$\nu$ is the parameter displayed in (\ref{52}) for that particular potential.
For a more general potential,
\beq
\nu = {1 \over n} {\del V \over {\del r^2}}\Biggr] _{\rm boundary}
\eeq
The volume element $d\mu$ for ${\bf CP}^k$ 
is normalized such that $\int d\mu =1$ and is given in local coordinates as
\beq
 d \mu = \epsilon^{i_1j_1i_2j_2\cdots i_kj_k}(\omega_K)_{i_1j_1} \cdots (\omega_K)_{i_{k}j_{k}} ~{d^{2k} x \over {(4 \pi)^k}} = k! \sqrt{\det \omega_K}~ {d^{2k} x \over {(2 \pi)^k}}
\label{dmuo}
\eeq
Since $\rho_0(r^2)$ is a step-function as in (\ref{step}), its derivative $\del\rho_0/ \del r^2$ produces a delta function with support at the boundary 
of the droplet. As a result the first two terms in (\ref{70}) are boundary terms.
The action $S_0$ in (\ref{70}) is a higher dimensional generalization of a chiral,  Wess-Zumino-Witten action, vectorially gauged
with respect to the time-independent background gauge field $\bar{A}$ \cite{nepomechie}. 
The third term is a WZW-term written as  an integral over a $(2k+1)$-dimensional region, corresponding to the droplet and time. The usual 3-form in the integrand of the WZW-term, $(G^\dagger d G)^3$, has now been augmented to the appropriate $(2k+1)$-form $(G^\dagger d G)^3 \wedge (\omega_K)^{k-1}$. Since the WZW-term is the integral of a locally exact form, the whole action $S_0$ should be considered as part of the edge action. 

The part of the action which depends on the external gauge field $A_\mu$ is given by
\beq
S_A= - {N \over N'}   \int dtd\mu ~\tr  \left[  \rho_0 *U^\dagger * \A *U \right]
\label{72}
\eeq
The large $n$ limit of $S_A$ was evaluated in \cite{KA2}. It contains a boundary contribution expressing the interaction between the matter field $G$ characterizing the edge excitations of the quantum Hall droplet and the external gauge field $A_\mu$ and a bulk contribution written solely in terms of the gauge field fluctuations $A_\mu$. Combining the large $n$ limits for $S_0$ and $S_A$, the total edge action is essentially a higher dimensional Wess-Zunimo-Witten action chirally gauged with respect to the external gauge field $A_\mu$ (
up to gauge invariant, completely $A$-dependent terms).
The bulk contribution is written in the form of Chern-Simons actions (when the fields $A_\mu$ are slowly varying with respect to the length scale set by $B$). 

To keep the expressions simple we first write down the results when $u^i =0$,
i.e., $\omega_K \rfloor dV =0$. 
The edge action is then
\beqar
S^{\rm edge}(u^i=0)
&=& {N \over {2nN'}} \Biggl[ \int \del_i \rho_0 (\omega_K^{-1})^{ij} G^{\dagger} (\del_0 + i A_0^LG -iG A_0^R)~G^{\dagger} (\del_j G +i A_j^L G -iG A_j^R) \nonumber \\
&&+{k\over {2 \pi}} \int \rho_0 \biggl[ -d \left( iA^L dG G^{\dagger} +i A^R G^{\dagger} dG + A^L G A^R G^{\dagger} \right)\nonumber\\
&&+{1 \over 3} \left( G^{\dagger} dG \right)^3 \biggr] \wedge \left( {\omega_K \over 2\pi} \right) ^{k-1}\Biggr] %\nonumber\\
~=S_{WZW} ( A^L = A + \bar{A}, A^R= \bar{A}) \label{81}
\eeqar
This is clearly 
a higher dimensional WZW action, gauged in a left-right asymmetric way. 
The full $S^{\rm edge}$ action, including the $u^i$ dependent terms,
is also a gauged WZW action; it is obtained from
(\ref{81}) by the following substitutions:
\beqar
&&\del_0 \rightarrow \del_{\tau}= \del_0 + u^k \del_k  \nonumber \\
&&A_0^L \rightarrow A^L_{\tau} = A_0 + u^k (A_k + \bar{A}_k), \qquad A_0^R \rightarrow A^R_{\tau} = u^k \bar{A}_k \nonumber \\
&&A_i^L = A_i + \bar{A}_i, \qquad A^R_i = \bar{A}_i 
\label{81b}
\eeqar
The derivative $\del_{\tau}$ is the convective derivative, well known in hydrodynamics. The appearance of $A_\tau$ is consistent with the gauging of the convective derivative.
One can explicitly verify that the $u$-dependent terms generated by
(\ref{81b}) from (\ref{81}) are gauge invariant. 

Because of the chiral gauging, $S^{\rm edge}$ (including the $u^i$-dependent terms) is not gauge invariant. Under a gauge transformation it changes by
\beq
\delta S^{\rm edge} = {Nk\over {4\pi nN'}} \int  d \rho_0 ~ \tr \left[  d (A + \bar{A}) \Lambda \right] \wedge \left( {\omega_K \over 2\pi} \right)^{k-1} \label{81a}
\eeq

The bulk contribution to the action is given by
\beqar
S^{\rm bulk}  &=&  -{N \over {N'}} \int dt d\mu~ \rho_0~ \tr \left(A_0 + u^k A_k \right) \nonumber \\
&&+ {kN \over {4 \pi n N'}} \int dt \rho_0\left[ \tr \biggl( (A+\bar{A}) d (A+ \bar{A}) + {2i \over 3} (A+\bar{A})^3 \biggr) \wedge \left( {\omega_K \over 2\pi} \right)^{k-1} \right.\nonumber \\
&&-{(k-1) \over 2\pi }\left. \tr \biggl[ \biggl( (A+\bar{A})d(A+ \bar{A}) +{2i \over 3} (A+\bar{A})^3  \biggr) dV \biggr] \wedge \left( {\omega_K \over 2\pi} \right)^{k-2}\right] \nonumber \\
&&+{N \over {2nN'}} \int dt d\mu ~\rho_0~ \tr \Big[    \nabla^i F_{ik}   + (k+1)  A_k
\Big] u^k
\label{77a}
\eeqar
The metric-dependent terms in the last line of (\ref{77a}) can be neglected compared to the rest of the terms. Written in terms of the dimensionful coordinates $\tilde{x}$, they get a prefactor
proportional to $1/(BR^2)$; they are small compared to the other terms
in the approximation where $R$ is large and the gradients of the external field are small compared to $B$. (The actions (\ref{81}) and (\ref{77a}) are related to the
K\"ahler-Chern-Simons and K\"ahler-WZW actions \cite{NS}.)

The $V$-dependent terms in $S^{\rm bulk}$ can be shown to be gauge invariant
for a spherically symmetric $\rho_0$ and $V$. The lack of gauge invariance
is entirely
due to the K\"ahler-Chern-Simons term in the second line of (\ref{77a}). 
The change in  $S^{\rm bulk}$ under a gauge transformation is
\beq
\delta S^{\rm bulk}  =  - {Nk \over {4 \pi nN'}} \int  d \rho_0  ~ \tr ~ \left[ d( A+\bar{A} ) \Lambda \right]  \wedge \left( {\omega_K \over 2\pi} \right)^{k-1} \label{86}
\eeq
Adding the gauge variations of the edge and bulk actions we find, as expected, that the total bosonic action $S$ is gauge invariant, $\delta S = \delta S^{\rm edge}  + \delta S^{\rm bulk} = 0$. (Anomaly cancellation for two-dimensional Hall effect is discussed in \cite{wen}.)

The phenomenon of anomaly cancellation is of course expected since gauge invariance is already built in the action $S$. The full bosonic action $S$ is, by construction, invariant under
\beq
\delta U  =  -i \lambda * U , \qquad \delta A_\mu  =  D_\mu \Lambda + i [A_\mu,~ \Lambda] 
\label{79a}
\eeq
via the induced $W_N$-transformation (\ref{11}).
This also implies the following gauge transformation for $G$,
\beq
\delta G~G^\dagger = -i \Lambda + \cdots 
\label{80}
\eeq
where the ellipsis indicates terms of higher order  in $1/n$.
This means that the large $n$ limit of the effective action 
$S = S^{\rm edge} + S_A^{\rm bulk}$
is automatically gauge invariant under (\ref{79a}) and (\ref{80}), guaranteeing the anomaly cancellation between the edge and bulk contribution.  

In the next section, we will outline a different derivation of the bulk action
by matrix techniques.
We shall see that the 
``topological" part of (\ref{77a}), where the last two terms are neglected, can be written, for $\rho_0={\bf 1}$, as a single $(2k+1)$-dimensional Chern-Simons term to all orders in $1/n$. 
In fact, 
with a little bit of algebra, and using $N/N' = n^k/k!$ for large $n$, the bulk action (\ref{77a}) can be brought to the form,
\beq
S_A^{\rm bulk} = S_{CS} (\tilde{A}), \qquad
\tilde{A} = \Big( A_0 +V,~ -a_i+\bar{A}_i+A_i \Big)\label{81c}
\eeq

The gauge fields in this equation are of the form $ A^a(T_a^T)$. Since only the last eigenvalue of the generator $T_{k^2+2k}$ contributes in the symbol, we may write the abelian part as $-a_i = -n a_{Ki} = a_{Ki} T^T_{k^2+2k} \sqrt{{ 2(k+1) \over k}}$. The combination $-a_i+\bar{A}_i+A_i$ can therefore be written as $a+ A$, where all components are expanded using $T^T$, $a$ being the full background field $\bar{A}^{k^2+2k} T^T_{k^2+2k} + \bar{A}^a T^T_a$. Notice that the antihermitian components are $i A^A T^T_A = -i A^A (-T^T_A) = -i A^A (T_A)_{\bar{R}}$, where the index $A$ denotes both the $U(1)$ and $SU(k)$ indices. 
The matrices $-T^T_A$ are the generators in the representation $\bar{R}$ conjugate to the representation $R$ of the $T_A$. We will use this in the next section. 
\vskip .05in
\noindent{$\underline{The ~nature ~of ~the ~edge ~states}$}
\vskip .05in
Turning now to the nature of the edge states,
this can be understood in terms of
the  field $G$ in (\ref{70}) \cite{KN2}. First consider the case of the background field being abelian, so
that one can write $G = e^{i\Phi}$ , where $\Phi$ is just one function, not a matrix.
The surface of the droplet is topologically $S^{2k-1}$. The action involves time-derivatives
of $\Phi$ and $D_\omega$ which is the derivative along an angular direction
on $S^{2k-1}$ which is $\omega_K$-conjugate to the radius of the droplet.
It is convenient to decompose $\Phi$ in terms of the eigenstates of $D_\omega$.
Since $S^{2k-1}/S^1 = {\bf CP}^{k-1}$, the surface of the droplet, other than the angular direction corresponding to $D_\omega$, will be ${\bf CP}^{k-1}$.
We see that for each eigenvalue
of $D_\omega$, $\Phi$ can be expanded in terms of the $\D$-functions
for $SU(k)$, with the eigenvalue for the right action of $R_{k^2-1}$ fixed to the
eigenvlaue for $D_\omega$. Explicitly, 
we can write
$\Phi = \sum_l \sum_{p,q|p-q=s}~c^{p,q}_m ~\D^{(p,q)}_{m;s} (h)$,
where $h \in SU(k)$, $\D^{(p,q)}_{m; s}(h)$
is the $\D$-function for the
irreducible representation of $SU(k)$ of the tensorial
type $T^q_p$ with $p$ symmetric lower indices, $q$ symmetric 
upper indices and
the contraction (or trace) of any $p$-type 
index with any $q$-type index must vanish.
The eigenvalue of $D_\omega$ is $s=p-q$, up to normalization factors. 
The right state $\vert s\ra$ in the $\D$-function denotes the unique
$SU(k-1)$-invariant state for this representation, with the given eigenvalue for
$R_{k^2-1}$.
Notice that $\D^{(p,q)}_{m;s}$ are similar to wave functions
of a reduced Landau problem on ${\bf CP}^{k-1}$.
More details of edge states can be found in \cite{ZH, eff-action, poly3}. The analysis of edge states on $S^4$ starting from
${\bf CP}^3$ can be found in \cite{KN2}.
For the case of a nonabelian background, one can carry out a similar analysis,
although the details are more involved. 

\section{The fuzzy space point of view}

We shall now return to the question of taking the large $N$ limit of a matrix action,
focusing on the fuzzy space-matrix model point of view.
The basic strategy has been to introduce a set of wave functions for the Hilbert space
${\cal H}_N$ and then the large $N$ limit can be defined using the symbols for the
matrices involved. But, as mentioned in the introduction, there are many ways to do this.
Since the
action we start with is a matrix action, there is, initially, no notion of space or spatial geometry.
The Hilbert space ${\cal H}_{\mathcal N}$ on which the matrices act as linear transformations can be taken, for example, as
arising from
 the quantization
of the phase space $S^2 = SU(2)/U(1)$, where the symplectic form is $\omega = -in \omega_K$, $\omega_K$ being the K\"ahler form on $S^2$. ($N$ will be a function of $n$.)
Taking $n$ large in this way defines a specific large $N$ limit.
Since $\omega$ is a background $U(1)$ field on $S^2$,  we could also consider  a deformation of this situation 
with, say, $\omega = -in \omega_K +F$,
where $F$ is topologically trivial (so that the dimension of ${\cal H}_N$ is not changed).
The wave functions to be used would now be modified and the
large $n$ limit, via the modified symbols, gives $S^2$ with a different choice of background field on it.
One could also consider ${\cal H}_N$ as the quantization of, say, 
${\bf CP}^k = SU(k+1)/U(k)$,
with a suitable choice of symplectic form (with the values of $N$ matching
the dimensions of a class of $SU(k+1)$ irreducible representations). This is what we did in the analysis for the dynamics of the quantum Hall droplet.
It is clear that there are many ways to take the large $N$ limit.
Even for the same geometry and topology for the phase space, the choice of symplectic form is not unique.
Once the dimension of the symplectic space has been chosen, these limits can be parametrized by the choice of background gauge fields. This is what we want to analyze,
particularly 
for the matrix action $S = i \int dt~ \Tr \rho_0(U^\dagger D_0 U)$, where
$D_0 = \del_0 +A_0$. (We will use antihermitian $A_0$ for simplicity of notation 
in this section. Also any potential $V$ can be included in $A_0$.) \footnote{
We do not use a hat to represent matrices or operators on ${\cal H}$ from now on
to avoid clutter in notation. Whether we mean the  matrix or the symbol should be clear from the context. $A_0$ plays now the role of $\hat{\A}$ of section 4.}

The result we find will be essentially identical, with some reinterpretation,
to the result for the quantum Hall system. However, keeping in mind the fuzzy geometry,
we want to take the point of view that the Hilbert space is the fundamental entity, with the smooth manifold being just a large $n$ simplification. 
It is, therefore, important to have a matrix version of the calculations
for extracting the large $n$ expansion.

The unitary transformation $U$ encodes the 
fluctuations of the chosen density matrix, or the edge states from the quantum Hall
point of view. Equivalently, it gives the boundary effects for dynamics in a subspace of a fuzzy space.
The bulk dynamics is not sensitive to $U$, and can be extracted by taking $\rho_0 = {\bf 1}$. Effectively, we are then seeking
the simplification of $S = i \int dt~\Tr D_0$ in the limit of large matrices.
This action is the one-dimensional Chern-Simons action for the matrix theory.

In the following, we shall choose a specific background and expand
the action around it. The final result is not
sensitive to the details of the background, except for the dimension and topology.
Therefore, we can choose a simple background, say, ${\bf CP}^k$ with
only the abelian field; thus $\omega = -in \omega_K$.
The gauge fields $A_0$, $A_i$, which can be abelian or nonabelian, will be expanded around this background; thus
the fields $A_0$, $A_i$ are actually functions on fuzzy ${\bf CP}^k$.

At this point, it is appropriate to clarify the relationship between the lowest Landau level and
fuzzy geometry in more specific terms.
The states of the lowest Landau level form an $N$-dimensional Hilbert space which we
identify as the space ${\cal H}_N$ needed for fuzzy ${\bf CP}^k$. Observables when restricted to the LLL are $(N\times N)$-matrices and these can be taken as functions on fuzzy 
${\bf CP}^k$. We can see that these are in correspondence with functions on smooth
${\bf CP}^k$. A basis for functions on smooth ${\bf CP}^k$ is of the form
$\{ \D^{R}_{m,w}(g)\}$ where $\vert w\ra$ is trivial under the action of
$U(k)\in SU(k+1)$, so that we get true functions on $SU(k+1)/U(k)$ and $R$ is any representation which contains such a state.
At the matrix level,
since the states are symmetric representations of $SU(k+1)$, a general matrix is of the form $X^{b_1 b_2 \cdots b_n}_{a_1 a_2 \cdots a_n}$ and transforms as
the product representation ${\bar J}\otimes J$. The reduction of this product will contain the singlet, the adjoint, and higher irreducible representations.
Thus, upon reducing
the product ${\bar J}\otimes J$, we can write a matrix $X$ in terms of a basis corresponding to the irreducible representations of $SU(k+1)$ as
\beq
X^{b_1 b_2 \cdots b_n}_{a_1 a_2 \cdots a_n} = \sum_{0\leq p \leq n}\sum_{\{\alpha\}}~
C^{A_1 A_2\cdots A_p} ~(T_{A_1 A_2\cdots A_p})^{b_1 b_2 \cdots b_n}_{a_1 a_2 \cdots a_n}
\label{diff10}
\eeq
The matrices $T_{A_1 A_2\cdots A_p}$ are obtained from products of the generators of $SU(k+1)$, namely $T_A$'s, with the condition that they are traceless
for any contraction of any of the $a_i$'s with any of the $b_j$'s. They form a complete basis 
at the  matrix level.
The symbol corresponding to the identity is the constant function on smooth ${\bf CP}^k$,
the symbol for $T_A$ will be of the form $\D^{(adj)}_{A,w}(g)$.
The symbols corresponding to 
$T_{A_1 A_2\cdots A_p}$ are $\D^{R}_{m,w}(g)$, for the appropriate representation $R$.
We see that the symbol corresponding to $X$ is a function on ${\bf CP}^k$, expandable in terms of a truncated set of basis functions since $p\leq n$.
As $n\rightarrow \infty$, ``functions'' on ${\cal H}_N$ tend to functions on 
${\bf CP}^k$. Further the star-product shows that the algebra of $Mat_N$ goes over to the commutative algebra of functions on ${\bf CP}^k$.

This argument is for wave functions of the LLL corresponding to an abelian background.
The wave functions for the LLL with a nonabelian background field are of the form $\D^J_{m, a'}(g)$, where the state $\vert J, a' , -n\ra$ transforms as the $J'$-representation of
$SU(k)$. This can be constructed in terms of a product of the abelian background
and another representation of $SU(k+1)$.
The state $\vert J, a' , -n\ra$ can be viewed as one set of states obtained by the reduction
of the product $\vert J_1 , -n\ra \otimes \vert J_2, a', 0 \ra$ for some representations
$J_1, J_2$ of $SU(k+1)$. In this way, matrices acting on
the product space of two $SU(k+1)$ representations can lead to the symbols
we obtained using the nonabelian wave functions.
This structure with two $SU(k+1)$ representations is what we expect for matter 
fields on ${\bf CP}^k$ which form an $SU(k+1)$ multiplet $J_2$, where one set
($J_1$ in our notation)
arises from the translations on the space. 
This is also the mathematical structure  relevant for the dynamics of a charged particle on a fuzzy space. (For example, for the fuzzy sphere, we find that two $SU(2)$ representations  are needed to define charged particle dynamics with a constant (monopole) background field \cite{KNP}.) In summary, we see that dynamics in the LLL for smooth ${\bf CP}^k$ can reproduce dynamics on fuzzy ${\bf CP}^k$.

There is also a description of
fuzzy ${\bf CP}^k$
directly in terms of embedding in ${\bf R}^{k^2 +2k}$, which will be useful in our discussion. For this we start with
$k^2 +2k$ hermitian matrices $X_A$ which are of dimension
$(N\times N)$,
where $N$ is of the form $(n+k)!/n!k!$ for some integer $n$.
The embedding conditions are then given by \cite{bal3, bal5}
\beqar
X_A X_A &=& {nk(n+k+1) \over 2(k+1)}~\equiv C_n\nonumber\\
d_{ABC} X_B X_C &=& (k-1) {(2n +k +1) \over 4 (k+1)}~ X_A ~\equiv \alpha_n ~X_A
\label{res21}
\eeqar
Consider the $SU(k+1)$-generators $T_A$ in the symmetric representation
of rank $n$. They may be written as
$T_A = a^\dagger_a (t_A)_{ab} a_b \equiv a^\dagger t_A a$, for bosonic annihilation-creation
operators $a_b, ~a^\dagger_a$, $a, b = 1, ..., (k+1)$.
By using completeness relations, one can easily prove that these obey representation-dependent identities which are identical to (\ref{res21}) with $T_A$ replacing $X_A$. In other words,
the matrices $T_A$ in the symmetric rank $n$ representation of $SU(k+1)$
give a solution of the embedding conditions (\ref{res21}) via $X_A = T_A
= a^\dagger t_A a$. This solution is obviously fuzzy ${\bf CP}^k$ since
functions of the $X$'s become general $N\times N$-matrices, acting on the symmetric rank $n$ representation of
$SU(k+1)$. In equation (\ref{res21}), $C_n$  is the quadratic Casimir operator
and $\alpha_n$ is another invariant related to the properties of the $d_{ABC}$-symbol.
We may also note that the conditions (\ref{res21}) can also be rewritten in terms of $-iT_A$ as
\beqar
(-iT_A) (-iT_A) &=& - C_n \nonumber\\
d_{ABC} (-iT_B ) (-iT_C) &=& -i \alpha_n  (-iT_A)\label{res22}
\eeqar

A general gauge field is introduced in the matrix language
by the prescription $D_A = -iT_A +A_A$. This will involve $k^2 +2k$ spatial
components for the gauge potential, which are obviously too many for
${\bf CP}^k$. Thus there are restrictions on $D_A$ which ensure that there are
only $2k$ spatial components for the potentials. These conditions may be taken as
the gauged version of the conditions (\ref{res22}), 
\beqar
D_A D_A &=& - C_n \nonumber\\
d_{ABC} D_B D_C &=& -i \alpha_n  D_A\label{GM3}
\eeqar
In other words, even after gauging, the derivatives obey the same embedding
conditions (\ref{res22}) as before gauging \cite{KNP}. (In the limit of a continuous manifold, there is some
redundancy in these conditions. While they are
sufficient for our purpose, whether they are necessary and sufficient
in the noncommutative case is not quite a settled issue.)

\section{The Chern-Simons action again}

We will now reconsider the simplification of the action
from a purely matrix point of view \cite{nair1}.
To carry out the expansion for large $N$, we write $A_0$, $A_A$ in terms of $(N\times N)$-blocks.
In other words, we can take ${\cal H}= {\cal H}_{N} \otimes {\cal H}_{2}$ so that the matrix elements of $A_A$, $i =0, 1, 2, ...$, may be written as
$A_{A pq} = \la p \vert A_A \vert q\ra = \la l ~a \vert A_A \vert r~ b\ra$,
$l, r = 1, 2, \cdots, N$, $a, b = 1, 2, \cdots, dim J_2$.
${\cal H}_{N}$ will carry an irreducible representation of $SU(k+1)$,
specifically the symmetric rank $n$ representation. ${\cal H}_2$
carries the representation $J_2$ of some compact Lie group.
(The action obtained in the previous sections is for the case when this group
is $SU(k+1)$ or a subgroup of it.)

We will consider the variation
of the matrix action $i\Tr D_0$ under a change of the background fields.
More generally, let
$K$ be a matrix acting on the Hilbert space ${\cal H}$. 
We may write
$K$ in an expansion in $D$'s as a sum of terms of the form
\beq
K = K(-iT)= K^{A_1 A_2 \cdots A_s} ~(-iT_{A_1}) (-iT_{A_2}) \cdots (-iT_{A_s})
\label{sph9}
\eeq
(Our results extend by linearity to sums of such terms, so it is sufficient to consider one such term,
for a fixed value of $s$.)
In the large $n$ limit, the $T$'s typically become the coordinates for the
space ${\cal M}$, in an embedding of ${\cal M}$ in ${\bf R}^d$ of suitable dimension $d$.
Since  $-i T_A = D_A - A_A$, it is possible to expand
$K$ in terms of the $D_a$'s which give the same basis on a background with
additional gauge fields $A_A$.
This can be done by writing
$K = K( D_A-A_A)$ and expanding in powers of $A$.
Since $A$ is not necessarily small, it is easier to
consider a perturbation around $D$, 
and calculate the variation
of $K$ by expanding $K(D-\delta D) - K(D)$ to linear order in $\delta D$.
By integrating this over $\delta D_A$ up to
$A_A$ we can obtain $K$. This will give an expression for $K$ in terms of $K(D)$.

The actual calculation will involve a number of steps:
\begin{enumerate}
\item We write the commutator of $D$'s as
$[D_A, D_B] = f_{ABC} D_C +F_{AB} \equiv \omega_{AB} +F_{AB} \equiv \Omega_{AB}$. 
This defines $\omega, \Omega$. When the gauge field fluctuations are zero, 
$\omega_{AB} = f_{ABC} D_C$ will become the symplectic form in the large $n$ limit.
We first define a matrix $N_{AC}$ which will play the role of an inverse to
$\Omega_{CB}$ when acting on functions of $D$'s which obey the embedding conditions (\ref{GM3}) and which tends to the inverse of
the symplectic form at large $n$.
\item We then write $\delta D$ in terms of $N_{AC}$, which will
generate a series which is naturally in powers of
$1/n$.
\item The variation of $K$ to first order can then be obtained in a suitable form.
In particular, we take $K=D_0$ to get the variation of the action,
$i\Tr (\delta D_0)$.
\item The next step will be to use the symbols to simplify the action. There is a correction to the definition of the symbol which must also be included.
\item The result can be compared to the variation of the Chern-Simons action to establish that the action does indeed become the Chern-Simons action.
\end{enumerate}
We shall now go over these steps, indicating briefly the basic mathematical
results involved.
\vskip .05in
\noindent{$\underline{An ~`inverse' ~ to ~ \Omega}$}
\vskip .05in
$N_{AB}$ is defined by the equations
\beqar
N_{AC} \Omega_{CB}  &=& \delta_{AB} + {\mathbb X}_{AB} + i {\mathbb Y}_{AB} \nonumber\\
\Omega_{BC} N_{CA} &=& \delta_{BA} + {\mathbb X}_{BA} + i {\mathbb Y}_{BA} 
\label{N11}
\eeqar
where
\beq
{\mathbb X}_{AB} = {D_B D_A \over B_n}, \qquad {\mathbb Y}_{AB}
= {1\over B_n}  \left( n +{\half} (k+1)\right) \left[ 
d_{ABC} D_C +i { \alpha_n\over 2} \delta_{AB}\right]
\label{N10b}
\eeq
and $B_n$ denotes the combination
$B_n = {1\over 4}{n (n+k+1)} + {1\over 16}{(k^2 -1)}$.
We also introduce the expression
\beq
N_{0AC} = {1\over B_n} \biggl[  f_{ACK} D_K + {1\over 4} (k-1) \delta_{AC}\biggr]
\label{N9}
\eeq
It obeys the equation $\bigl( N_0 \omega )_{AB} 
=\delta_{AB} + {\mathbb X}_{AB} + i {\mathbb Y}_{AB} + {\mathbb R}_{AB}$
with
\beqar
B_n ~{\mathbb R}_{AB} &=& f_{ACK} D_K f_{CBL}  D_L - D_B D_A - {C_n \over k} \delta_{AB}
- {i \over 2}{(2n +k +1)} d_{ABC} D_C \nonumber\\
&&+ {1\over 4}{(k-1)} f_{ABC} D_C \label{N1}
\eeqar

The matrices ${\mathbb X}$ and ${\mathbb Y}$ are of order $1$ at large $n$; 
${\mathbb R}$ is naively of the same order, but it is actually of lower order
due to algebraic identities on $T_A$.
A solution for $N_{AB}$ can be obtained as a series by writing $N = N_0 + N_1 + N_2+  \cdots$, and matching terms of the same order in powers of $n$ in (\ref{N11}).
The first few terms are given by
\beqar
N_{AB} &=& N_{0AB} - ({\mathbb R} N_0 )_{AB}  - (N_0 F N_0 )_{AB} 
+ ( {\mathbb R}({\mathbb X} + i {\mathbb Y} ) N_0 )_{AB}
+( N_0 F({\mathbb X} + i {\mathbb Y} ) N_0 )_{AB}\nonumber\\
&&+ (N_0 F N_0 F N_0)_{AB} +\cdots
\label{N12}
\eeqar
The embedding conditions (\ref{GM3}) are crucial in verifying 
that this is a solution to 
(\ref{N11}).
Terms containing powers of ${\mathbb X}, ~{\mathbb Y}$, such as
${\mathbb R} {\mathbb X}  N_0$, ${\mathbb R}{\mathbb Y} N_{0}$, are seemingly of the same order ${\mathbb R}N_0$ , since ${\mathbb X}, ~{\mathbb Y}$ are of order $1$.
But they are actually down by a power of $n$ due to the embedding conditions.
The series (\ref{N12}), therefore, is appropriate at large $n$.
\vskip .05in
\noindent{$\underline{Expression ~for ~\delta D}$}
\vskip .05in
Using $N_{AB}$, we can express the variation of $D$ in a form suitable for
a series expansion in $1/n$.
Multiplying equations (\ref{N11}) by $\delta D_A$,
using $\delta D_A {\mathbb Y}_{AB} + {\mathbb Y}_{BA} \delta D_A =0$,
which also follows from the embedding conditions,
and rearranging terms one can show that
\beqar
\delta D_B &=&  {1\over 2}\bigl( \xi_C [D_C , D_B ] +
[D_B , D_C ]{\tilde\xi}_C \bigr)
-{1\over 4B_n}[ \delta D \cdot D - D\cdot\delta D , D_B ]\nonumber\\
\xi_C &=&\delta D_A \left( N_{AC} +{\delta_{AC}\over B_n}\right) , \qquad
{\tilde \xi}_{C}  = \left( N_{CA} +{\delta_{CA}\over B_n}\right)\delta D_A
\label{mat4}
\eeqar

\noindent{$\underline{Variation ~of ~ K}$}
\vskip .05in
We now turn to the matrix function
$K = K^{A_1 A_2 \cdots A_s} D_{A_1} D_{A_2} \cdots D_{A_s}$. Here we can
take, without loss of generality, the coefficients $K^{A_1 A_2 \cdots A_s}$ to be symmetric in all indices. (Any antisymmetric pair may be reduced to a single $D$ and $F$; $F$
itself may be re-expanded in terms of $D$'s, to bring it to this form.)
Taking the variation of $K$ under $D\rightarrow D -\delta D$, and rearranging terms keeping in mind this symmetry, 
we can bring $\delta K$ to the form
\beq
\Tr ~(\delta K) = -{1\over 2} \Tr \Bigl[ \delta D_A N_{AB} [D_B, K] 
- [D_B, K] N_{BA} \delta D_A + {\Or}(1/n^3)\Bigr]\label{mat13}
\eeq
This gives the change in $\Tr~ K$ to
order $1/n^2$, as $n$ becomes large.
\vskip .05in
\noindent{$\underline{Variation ~of ~ \Tr (D_0)}$}
\vskip .05in
Equation (\ref{mat13}) can be used to work out the
expansion of the action, at least to order $1/n^2$, taking
$K= D_0$. The terms in $N_{ab}$, from equation (\ref{N12}), which can contribute to this
order, are
\beq
N_{AB} = \omega^{-1}_{AB}   + {(k-1) \over 4 B_n} \delta_{AB}  
-{\omega^{-1}_{AC}~F_{CD} ~\omega^{-1}_{DB} } - {\mathbb R}_{AC} 
~\omega^{-1}_{CB}
~+{\Or}(1/n^3)
\label{mat14}
\eeq
where $\omega^{-1}_{AB}  = {f_{ABC} D_C / B_n}$
(The notation anticipates the fact that this matrix will become
the inverse of $\omega$ in the large $n$ limit. But,
at this stage $\omega^{-1}_{AB} $ is still a matrix.)
The variation of the action, up to order $1/n^2$, is obtained from (\ref{mat13}),
(\ref{mat14}) as
\beqar
\Tr~(\delta D_0) &=&
\delta D_0^{(1)} + \delta D_0^{(2)} +\cdots\nonumber\\
\delta D_0^{(1)}&=& - {1\over 2} \Bigl( \delta A_A \omega^{-1}_{AB}  F_{B 0}
+ F_{B 0} \omega^{-1}_{AB}  \delta A_A\Bigr)\label{mat16}\\
\delta D_0^{(2)} &=& {1\over 2} \Bigl(\delta A_A \omega^{-1}_{AB}  F_{BC}
\omega^{-1}_{CD} F_{D 0}
+ F_{D 0} \omega^{-1}_{AB}  F_{BC} \omega^{-1}_{CD} \delta A_A
\Bigr)\nonumber
\eeqar
\vskip .05in
\noindent{$\underline{Using ~symbols}$}
\vskip .05in
What is left is to simplify the expansion (\ref{mat16}), which is still in matrix terms,
in terms of the symbols as an integral over
${\bf CP}^k$ with a trace over the remaining (small, $a,b$-type) matrix labels.
For this, we can bring the $\omega^{-1}_{AB} $
to the left end by the cyclicity of the trace, and then 
replace it by
\beq
\omega^{-1}_{AB}  ={1\over B_n}\Biggl[  -i {nk \over \sqrt{2k(k+1)}} f_{ABC} S_{C,k^2+2k} 
+ {i\over 2} f_{ABC} S_{C-i} R_{+i}
+ f_{ABC} S_{C\alpha} A_\alpha\Biggr]
\label{mat17}
\eeq
Here we have used the standard rule for simplifying 
the symbol of $T_A X$ and also the fact the symbol of the gauge field may be written
as $A_C = S_{C\alpha}A_\alpha$ where the summation is over $\alpha = 1$ to $2k$.
The fact that the symbol of $A_C$ has this restricted form is due to the constraints
(\ref{GM3}).
The inverse of $\omega$, in the limit of the continuous manifold,
is given in the coordinate basis as
\beq
\omega^{-1ij}= -i {nk \over B_n\sqrt{2k(k+1)}}~ f^{\alpha\beta, k^2+2k} (E^{-1})^i _\alpha
(E^{-1})^j_\beta
\label{mat17a}
\eeq
where $E$'s are the frame fields for the metric on ${\bf CP}^k$.
This can be used to simplify the first term on the right hand side of (\ref{mat17})
as $\omega^{-1ij} E^\alpha_i E^\beta_j S_{A\alpha} S_{B\beta}$.
\vskip .05in
\noindent{$\underline{Change ~ in ~symbol}$}
\vskip .05in
There is one more
correction which we must take account of. The symbol was defined using wave functions
with the gauge field fluctuations equal to zero.
As the potential is changed, the definition of the symbol also changes. This change
can be calculated as
\beq
(K) = (K)_0 - {1\over 4}  \bigl[ (\omega^{-1}_{AB} F_{AB} + F_{AB} \omega^{-1}_{AB})~K\bigr]
~+ \cdots
\label{mat20}
\eeq
This is needed to simplify the symbol for $\delta D_0^{(1)}$,
for
which $K = - {1\over 2}( \delta A_A \omega^{-1}_{AB} F_{B 0}
+ F_{B 0} \omega^{-1}_{AB} \delta A_A )$.
\vskip .05in
\noindent{$\underline{The ~variation ~ of ~the ~action}$}
\vskip .05in
Taking account of
these observations, the evaluation of the action is straightforward, although somewhat tedious.
The result, to order $1/n^2$, is obtained as
\beqar
\int dt \Tr (\delta D_0 ) &=& N \int dt d\mu ({\bf CP}^k)
\Biggl[-  \omega^{-1ij} \tr ( \delta A_i F_{j 0})\nonumber\\
&&+ {1\over 2} \left( \omega^{-1im} \omega^{-1nj}
+  {1\over 2} \omega^{-1ij}\omega^{-1mn} \right)~\tr 
\Bigl[(\delta A_i F_{j 0} + F_{j 0} \delta A_{i }) F_{mn}\Bigr]
\nonumber\\
&&-{1\over n}~ \omega^{-1mn}
\tr \Bigl[ \delta A_m ~\left(-D^2 +(k+1)\right)~F_{n 0} \Bigr] +{\Or}(1/n^3)\Biggr]
\label{mat29}
\eeqar
In this expression, the field components are in the coordinate basis.
\vskip .05in
\noindent{$\underline{Relation~ to ~the~ Chern-Simons ~action}$}
\vskip .05in
Equation (\ref{mat29}) can be expressed in terms of the Chern-Simons form.
The variation of the $(2k+1)$-dimensional Chern-Simons term is given by
\beq
\delta S = {i^{k+1} \over (2\pi )^{k} k!} \int \tr \bigl( \delta A F^k \bigr)
\label{mat30}
\eeq
Replacing $F$ by $\omega +F$ and expanding, we get
\beq
\delta S = {i^{k+1} \over (2\pi )^{k} k!} \int \tr \Bigl( \omega^k \delta A  + k \omega^{k-1} 
\delta A F + \half k (k-1)  \omega^{k-2} \delta A F^2 +\cdots \Bigr)
\label{mat31}
\eeq
Since $\omega$ is proportional to the K\"ahler form, this can be simplified and written in terms 
of the standard volume measure $d\mu$ for ${\bf CP}^k$.
As an example, we note that the second term can be written as
\beq
{i^{k+1} \over (2\pi )^{k} k!}~k \omega^{k-1} \tr (\delta A_m F_{n 0} )~ dx^m  dx^n  dt
=i {n^k \over k!} \int dt d\mu  ~ \Bigl[ - \omega^{-1mn} \tr (\delta A_m F_{n 0})\Bigr]
\label{mat33}
\eeq
Rewriting the other terms similarly, and comparing with (\ref{mat29}), we find
\beqar
i \int \! dt~\Tr (\delta D_0) &=& {N k! \over n^k}~\delta S_{CS} 
-i{N\over n} \int dt d\mu~ \omega^{-1mn}~
\tr \Bigl[ \delta A_m \left(-D^2 +(k+1)\right)F_{n 0} \Bigr]+\cdots\nonumber\\
&\approx& \delta S_{CS} -i{N\over n} \int dt d\mu ~ \omega^{-1mn}
\tr \Bigl[ \delta A_m \left(-D^2 +(k+1)\right) F_{n 0} \Bigr] \! +{\Or}(1/n^3)
\label{mat37}
\eeqar
In this expression for $i\Tr (D_0)$ we have also included a term
\beq
i \Tr \delta_f A_0 = {N k! \over n^k} {i^{k+1} \over (2\pi )^{k} k!} ~\int
\omega^k ~\tr (\delta A_0)
\label{mat36}
\eeq
The reason is that (\ref{mat29}) only gives the variation of $i\Tr (D_0)$ due to
the change in the spatial components of $A$,
namely, under $A_i \rightarrow A_i +\delta A_i$.
A change in the functional form of $A_0$ is also possible; the variation of $i \Tr (D_0)$ due to this is (\ref{mat36}) and should be included to obtain the general variation.

The first term on the right hand side of (\ref{mat37}) will integrate to give the Chern-Simons form. The second term is due to the higher terms, terms involving derivatives, in the star-product. 
Let $S_{*CS}$ denote the Chern-Simons term defined with star-products used for the products of
fields and their derivatives occurring in it.
The integrated version of (\ref{mat37}) is then
\beq
i\int dt~ \Tr (D_0) \approx S_{*CS}~+~\cdots
\label{mat39}
\eeq
It is worth emphasizing that
the gauge potentials in this Chern-Simons action are the full potentials
$a +A$, where $a$ is the background value corresponding to the symplectic form
and $A$ is the additional potential or gauge field fluctuation.

The second term in (\ref{mat37}) which arises from higher terms in the star-product
also agrees with the result (\ref{77a}), if we write $A_0 =i V$; some partial integrations are also necessary. As argued after (\ref{77a}), by rescaling the coordinates $x \rightarrow \tilde{x} = Rx$, this term is seen to be small if the radius is large and the gradients of fields are small compared to the value of the background field.
The Chern-Simons form is unaffected by this scaling. In this approximation, the 
leading term of the action is given by
\beq
i\int dt~ \Tr (D_0) \approx S_{CS}~+~\cdots
\label{mat43}
\eeq
This result shows that the expansion of
$\Tr (D_0)$ around different backgrounds can be approximated, in the large $n$ limit
and for small gradients for the field strengths,
by the Chern-Simons form, with $A$ replaced by $a+A$, $a$ being the desired background
potential. Notice that any reference to the metric and other geometrical properties of
${\bf CP}^k$ has disappeared in this result.

Strictly speaking, our calculation has explicitly verified this result (\ref{mat43}) only up to
order $1/n^2$, or, equivalently, up to the term involving the $5$-dimensional
Chern-Simons form for $A$. To this order, we do get $S_{CS}(a+A)$.
But we can see that the result (\ref{mat43}) holds with the full Chern-Simons term.
This is because, the final expression, whatever it is, should be a functional of only the combination  $a+A$, since the separation between the background and fluctuation is arbitrary. Also
it should have the correct gauge invariance property and it should agree with $S_{CS}(a+A)$ when expanded up to the term with the
$5$-dimensional Chern-Simons form for $A$. The only such term, apart from the ambiguity of higher
gradients of fields, is $S_{CS}(a+A)$. 
Thus the result (\ref{mat43}) holds in general, in a gradient expansion at large $R$.

\section{Towards a matrix theory of gravity}

The basic mathematical result we have can be applied to gravity on fuzzy spaces
\cite{nair2}.
As mentioned in the introduction, one may take the minimalist point of view that fuzzy spaces are just another regularization. Then, for ordinary field theories it would not be anything special, but it is still very attractive for gravity since symmetries can be preserved.
One may also consider fuzzy space as fundamental, continuous space being a large $N$ approximation. Formulating gravity on fuzzy spaces is, in either case, an interesting 
problem. (For earlier formulations of gravity on fuzzy spaces, see
\cite{connes2, abe}.)

To see how gravity arises naturally, recall that the background fields we have considered are valued in the Lie algebra of $U(k)$ for 
${\bf CP}^k= SU(k+1)/U(k)$. This is part of the isometry group $SU(k+1)$ of ${\bf CP}^k$. Gauging of the isometry group
introduces gravity, so we may interpret the gauge fields as gravitational fields.
This is the basic point of contact. We
shall now present an argument on how our results may be adapted for
describing gravity.

The setting for the problem is the finite-dimensional Hilbert space ${\cal H}$, which we may take to be
split into a matter part ${\cal H}_m$ and a space part ${\cal H}_s$, the latter leading to the spacetime at large $n$. The action for the evolution of states is given by the action
\beq
S = i \int dt~ \Tr \rho_0 (U^\dagger D_0 U)
\label{grav1}
\eeq
where $D_0 = \del_0 + A_0$. The Hamiltonian as a matrix on the Hilbert space
is $H = -i A_0$. We know that (\ref{grav1}) is the most general equation for evolution of states for matter, the specific characteristics of the matter system being encoded in the choice of
$H$ and other observables. The only natural choice
for the space part is that the same action should apply to evolution within ${\cal H}_s$.
To see how this can be implemented, represent a general state in ${\cal H}$ as
$\vert A,r\ra$, where the
labels $A, B,$ etc., pertain to the degrees of freedom of space
and the labels $r, s,$ etc., describe the matter system of interest. 
For the operator $D_0$, we introduce the splitting
\beq
\la A, r\vert D_0 \vert B, s\ra
=  \delta_{rs} ~\la A \vert D^{(s)}_{0}\vert B\ra ~+~ \la A, r\vert D^{(m)}_0
\vert B, s\ra
\label{grav3}
\eeq
The part of $D_0$ which is proportional to the identity in ${\cal H}_m$ is designated
as $D^{(s)}_0$ and the remainder as $D^{(m)}_0$. The latter includes effects
of coupling the matter system of interest to the spatial degrees of freedom.
We also take the density matrix to have the form
\beq
\la A, r\vert \rho_0 \vert B, s\ra = \delta_{AB} ~\la r \vert \rho_0 \vert s\ra
\label{grav4}
\eeq
Notice that we take $\rho_0$ to have maximal rank in ${\cal H}_s$;
if the rank is less than maximal, it would
mean that the dynamics does not cover all of space. This is why we make the choice (\ref{grav4}).

While $A_0$ (or $H$) specifies the choice of matter system, for spacetime,  the geometry is
not a priori determined, therefore $D^{(s)}$ should be regarded as an arbitrary matrix
in (\ref{grav1}).
Thus we take the action (\ref{grav1})
as the action for the theory, including gravity, where $U$ and $D^{(s)}_0$ are regarded as 
quantities to be varied. $D^{(m)}_0$ is to be regarded as a given operator, specifying the matter system of interest. 

We can get different large $n$ limits for the action depending on
what backgrounds we choose. Extremization of the action can be used to determine
the best background. If we ignore all matter degrees of freedom as a first
approximation, the action becomes
\beq
S \approx i \int dt~ \Tr (D^{(s)}_0)
\label{grav4}
\eeq
For the case of ${\bf CP}^k$ with a nonabelian background, the
wave functions were of the form $\D^J_{m, a'}(g)$.
As mentioned in section 8, the state $\vert J, a' , -n\ra$
may be taken as one set of states obtained by the reduction
of the product $\vert J_1 , -n\ra \otimes \vert J_2, a', 0\ra$ for some representations
$J_1, J_2$ of $SU(k+1)$. (We will take $J_2$ to be the fundamental representation of
$SU(k+1)$ for simplicity.)
Therefore, at the matrix level we split the states as ${\cal H}_s = {\cal H}_N\otimes
{\cal H}_2$, where the components are of dimensions $N= dim J_1$ and 
$dim J_2= k+1$ respectively. Correspondingly, we write $D_0$ as
$D_{0pq} = \la p \vert ~D_0 ~\vert q\ra = \la l ~a \vert ~ D_0 ~\vert r~ b\ra$,
$l, r = 1, 2, \cdots, N$, $a, b = 1, 2, \cdots, k+1$. The matrix structure
for the indices $l, r$ will be converted to the symbol, and we carry out a large
$N$-expansion.
The result is $(2k+1)$-dimensional Chern-Simons theory. For the gauge fields, in general, there will
be a $U(1)$ component as well. Thus the gauge group is $U(k+1)$.
The conclusion of this argument is that, for a fuzzy space, we should expect
Chern-Simons gravity \cite{zanelli}.
(In this context, it is fascinating that
 there are indications of Chern-Simons gravity in the context of
M-theory \cite{horava}; we expect that the present analysis can be related to
a matrix version of some of the
considerations in these references.)

The simplest example along these lines would be $k=3$, which gives
a $U(4)$-Chern-Simons theory on a seven-dimensional space.
We take this space to be of the form $S^2 \times M^5$ and the gauge field
strength as $-il \omega_K + F$ where $\omega_K$ is the K\"ahler form
on $S^2$, $l$ is an integer and $F$ is in the $SU(4)$ Lie algebra.
The action is then given by
\beq
S = -i {l \over 24 \pi^2} \int \tr\left( A ~dA~ dA + {3\over 2} A^3~ dA +{3\over 5} A^5\right)
\label{grav7}
\eeq
Since $SU(4)$ is locally isomorphic to $O(6)$, we see that this is appropriate for
Euclidean gravity in five dimensions.
In fact,  the $SU(4)$ potential can be written as
$A = P^\alpha E^\alpha_i dx^i + \half  J^{\alpha\beta} \omega^{\alpha\beta}_i dx^i$
where $J_{\alpha\beta}$ are the generators of $O(5) \subset O(6)$ and
$P^\alpha$ are a basis for the complement of $\underline{O(5)}$ in $\underline{O(6)}$.
$E^\alpha$ are the frame fields and $\omega^{\alpha\beta}$ is
the spin connection.
The equations of motion give $F =0$ and has the solution
$A = g^{-1} d g$ with $g\in O(6)$. This is with no matter.)
This space is $O(6)/O(5) =S^5$ which is the Euclidean version of de Sitter space.
It is given in a basis where the
cosmological constant $\Lambda$ has been scaled out; it may be introduced
by the replacement
$E^\alpha \rightarrow \sqrt{\Lambda} ~E^\alpha$. 

There is also a neat reduction of this to four dimensions \cite{CMM}. 
This is achieved by the additional restrictions $E^5_5 =1$, $\omega^{5\beta} =0$, 
$\omega^{\alpha\beta}_5 =0$,
for $\alpha, \beta = 1, ... , 4$. The fifth dimension is taken as a circle of, say, unit radius.
(This restriction, as well as the choice of $S^2$ with the $U(1)$ field proportional to the K\"ahler form, can be interpreted as particular compactifications.)
In this case, the action becomes
\beqar
S &=& {l \Lambda \over 64\pi} \int \left( E^\alpha E^\beta R^{\gamma\delta} - {\Lambda\over 2}  E^\alpha E^\beta E^\gamma E^\delta \right) \epsilon_{\alpha\beta\gamma\delta}\nonumber\\
&=&{l \Lambda \over 16\pi} \int d^4x \sqrt{g} ~( R - 3 \Lambda ) 
\label{grav14}
\eeqar
We get the Einstein-Hilbert action with a cosmological constant.

There are many issues, such as getting Minkowski signature, generalizing to other dimensions
and incorporating matter in a detailed way, which are not yet clear.
Nevertheless, this nexus of Hall effect, fuzzy spaces and gravity is very suggestive and intriguing.

\section{Discussion, Outlook}

We shall begin with a synopsis of the basic result of our analysis.
We have a finite-dimensional Hilbert space ${\cal H}$ with fields and other observables 
realized as linear transformations or matrices acting on this space.
When the dimension of the matrices is large, we can simplify matrix products, the action, etc.,
using symbols for the matrices and star-products. The symbols are defined in terms of a set of wave functions.
These wave functions are based on a continuous smooth symplectic space ${\cal M}$,
of dimension, say, $2k$,
with a set of gauge fields defined on it. (We may think of
the Hilbert space ${\cal H}$ as providing a fuzzy version of ${\cal M}$.) The wave functions 
characterize the space ${\cal M}$
and the fields on it, as far as observables are concerned. 
The large $N$ limits of matrices are thus parametrized by ${\cal M}$ and the gauge fields.
Alternatively, we may think of the Hilbert space as the set of lowest Landau levels
for quantum Hall effect on ${\cal M}$ and the gauge fields as external fields to which the fermions couple.
In either case, the basic action we have analyzed is of the form
\beq
S = i \int  dt~ \Tr ({\hat\rho}_0 ~{\hat U}^\dagger {\hat D}_0 {\hat U})
\label{disc1}
\eeq
where ${\hat \rho}_0$ characterizes a fiducial or initial state.
Our basic result is then the following. 
\begin{itemize}
\item In the large $N$ limit, ${\hat\rho}_0$ describes a droplet of a subspace of
${\cal M}$. The simiplification of the action yields a
a bulk action and a boundary action.
\item The bulk action is given by the $(2k+1)$-dimensional Chern-Simons action,
when the gradients of the gauge fields are small.
This action, although obtained by expansion around a chosen background, is not
sensitive to the geometry of the space.
\item The boundary action describes the fluctuations of the boundary of the droplet,
or equivalently, the large $N$ limit of embeddings of a fuzzy sphere in the fuzzy version of ${\cal M}$.
It is given by a chiral gauged higher dimensional generalization of the WZW action.
\item The bulk and boundary actions are not separately gauge-invariant, but
the total action is, with the gauge anomalies canceling between the two.
\end{itemize}

Perhaps the most interesting conclusion which emerges from this analysis is the possibility of describing a number of higher dimensional theories as matrix models. For example, the $(2+1)$-dimensional Chern-Simons and the $2$-dimensional WZW theories help to define conformal field theories in two dimensions. One can introduce  a matrix model for them, as a specific
large $N$ limit of (\ref{disc1}). But such a matrix model can also lead to higher dimensional Chern-Simons and WZW models, as a different way of taking the large $N$ limit. This suggests a way of generalizing conformal field theories to higher dimensions. In this context, the exploration of some of the well known features of WZW models such as symmetry structures and current algebra would be very interesting. (This is also closely related to the 
K\"ahler-Chern-Simons and K\"ahler-WZW models \cite{NS}.)

Noncommutative Chern-Simons theories have been extensively investigated
over the last few years \cite{poly1}.
Properties of such theories on flat noncommutative spaces are fairly well understood by now. They have also been formulated on some $(2+1)$-dimensional fuzzy spaces,
but a general formulation on fuzzy spaces has not yet been possible \cite{poly2}.
These theories are
matrix versions of the Chern-Simons theory characterized by the choice of
the fuzzy space and the gauge group and
give the usual Chern-Simons theory at large $N$, just as
the action (\ref{disc1}) does.
It is perfectly sensible to study these matrix models and (\ref{disc1})
as different theories, but
if we only ask for a matrix theory whose commutative limit
gives the Chern-Simons theory, then the action (\ref{disc1}) is a good choice.
It has also the advantage that it can be easily formulated on any fuzzy space
and can give Chern-Simons theories on smooth spaces of any dimension and with any gauge group, depending on how the large $N$ limit is taken.
The matrix version (\ref{disc1}) may thus be considered as a `universal' 
Chern-Simons theory \cite{nair1}.

Bosonization in higher dimensions is another closely related topic \cite{boso1, poly4}.
The phase space for fermions in $k$ dimensions is $2k$-dimensional.
Semiclassically, each quantum state corresponds to
a certain phase volume. Thus, by the exclusion principle, a collection of a large number of
fermions is an incompressible droplet in the phase space.
Deformations of the droplet give a bosonic description of the dynamics of this collection of fermions.
Thus the matrix action (\ref{disc1}) can also be used for phase space bosonization.
The large $N$ result is evidently the generalized WZW theory.
Related approaches in formulating higher dimensional phase space bosonization in terms of a noncommutative field theory have been explored in \cite{poly3, poly4}.

There is also an evident connection to fluid dynamics; the edge dynamics of the droplet
is that of an incompressible droplet of fluid. The additional gauge fields allow for nonzero compressibility. The droplet can be viewed as the embedding of a $2k$-brane in ${\cal M}$.
Therefore one should  be able to relate these ideas to the descriptions of fluids in the brane language \cite{fluids}.
It is also related to the noncommutative description of the quantum Hall effect proposed in \cite{suss}.

As mentioned in the last section, gravity on a fuzzy space may be the context in which these results can be most fruitful. This story is far from complete, there are many issues related to
the Minkowski signature, incorporation of matter, etc. which need to be clarified. 
Also, as mentioned earlier, there are suggestions that
Chern-Simons gravity can provide an effective description of $M$-theory \cite{horava}.
A matrix description via (\ref{disc1}) is an attractive possibility that needs to be explored further.
It is also suggestive that
quantum Hall droplets appear in the dual field theories for many gravitational backgrounds
\cite{beren}.

Another interesting line of development, which we have not discussed, is the supersymmetric version of quantum Hall effect \cite{hasebe}. The bosonic partners of fermions do not have to form an incompressible droplet since there is no exclusion principle for them.
Nevertheless, it is possible to obtain supersymmetric droplets and study their 
properties.
This can be applicable in the context of supersymmetric brane dynamics, supergravity, etc.
\vskip .1in
This work was supported in part by the National Science Foundation grants
PHY-0457304 and 
PHY-0244873 and by PSC-CUNY grants.


\section*{References}
\begin{thebibliography}{199}

\bibitem{bal1} For a recent review of fuzzy spaces and theories defined on them, see
Balachandran A P 2002 {\it Pramana} {\bf 59} 359 \\
Balachandran A P and  Kurkcuoglu S 2004  \IJMPA ~{\bf 19}  3395\\
Balachandran A P, Kurkcuoglu S and Vaidya S 2005 {\it Lectures on Fuzzy and Fuzzy SUSY Physics} {\it Preprint} hep-th/0511114

\bibitem{bal2} Balachandran A P, 
Govindarajan T R and Ydri B 2000 \MPLA~ {\bf 15} 1279 \\
Balachandran A P, 
Govindarajan T R and Ydri B 2000 {\it Preprint}
hep-th/0006216 \\
Aoki H, Iso S and Nagao K 2003 \PRD~{\bf 67} 065018

\bibitem{connes1} Connes A 1994 {\it
Nocommutative Geometry}  (Academic Press)\\
Madore J 1995 {\it An Introduction to Noncommutative Geometry and its Physical
Applications},  LMS Lecture Notes 206 \\
Landi G 1997
{\it An Introduction to Noncommutative Spaces and their
Geometry}, Lecture
Notes in Physics, Monographs m51 (Springer-Verlag)

\bibitem{taylor} For a review of the matrix version of M-theory and its solutions, see,
Taylor W IV 2001
\RMP ~{\bf 73}  419 

\bibitem{DN} For a review of field theories on noncommutative spaces, see,
Douglas M R and Nekrasov N A 2001 \RMP ~{\bf 73}  977

\bibitem{connes2} Chamseddine A, Felder G and
 Fr\"ohlich J 1993
\CMP ~{\bf 155} 205\\
 Kalau W and 
 Walze M 1995
\JGP ~{\bf 16} 327\\
Kastler D 1995 \CMP ~{\bf 166} 633\\
Chamseddine A and Connes A 1997
\CMP ~{\bf 186} 731\\
Madore J 1997 {\it Preprint} gr-qc/9709002\\
Ho P-M 1997 \IJMPA~ {\bf 12} 923\\
Moffat J 2000 \PLB~{\bf 491} 345 \\
Banados M {\it et al} 2001 \PRD~{\bf 64} 084012\\
Cacciatori  S {\it et al} 2002 \CQG ~{\bf 19} 
4029\\
Viet N A and Wali K C 2003 \PRD~{\bf 67} 124029\\
Chamseddine A 2003 \JMP~ {\bf 44} 2534\\
Castro C 2004 \GRG~{\bf 36} 2605 \\
Aschieri P {\it et al} 2005 \CQG~{\bf 22} 3511 \\
 Aschieri P {\it et al} 2006 \CQG~{\bf 23} 1883\\
Balachandran A P, Govindarajan T R, 
Gupta K S and Kurkcuoglu S 2006 {\it Preprint} hep-th/0602265


\bibitem{dewit} de Wit B, Hoppe J and Nicolai H 1988 \NPB~ {\bf 305} 545

\bibitem{mtheory}
Banks T, Fischler W, Shenker S and
Susskind S 1997
\PRD~{\bf 55} 5112\\
Ishibashi N, 
Kawai H, Kitazawa Y and Tsuchiya A 1997 \NPB~{\bf 498} 467

\bibitem{mald} There are many papers on this; the paper Berenstein D, Maldacena J M and Nastase H 2002 \JHEP ~{\bf 0204} 013 and citations to it can give a perspective on matrices and branes.


\bibitem{KNR} Karabali D, Nair V P and Randjbar-Daemi S 2004 in 
Shifman M, Vainshtein A and Wheater J (eds.) {\it From  Fields to Strings: Circumnavigating Theoretical Physics,
Ian Kogan Memorial Collection}
(World Scientific) p~831-876

\bibitem{2dqhe} For a recent review on QHE, see Girvin S 2000
in Comtet A {\it et al} (eds.)  {\it Topological Aspects of Low Dimensional Systems}
(Springer-Verlag, Berlin), and references therein ({\it Preprint} cond-mat/9906454)

\bibitem{ZH} Zhang S C and Hu J P 2001 {\it Science} {\bf 294}  823 \\
Hu J P  and  Zhang S C 2001 {\it Preprint} cond-mat/0112432

\bibitem{KN1} Karabali D and Nair V P 2002 \NPB~ {\bf 641}  533

\bibitem{KN2} Karabali D and Nair V P 2004 \NPB~ {\bf 679}  427 \\
Karabali D and Nair V P 2004 \NPB~ {\bf 697}  513

\bibitem{everyone} Fabinger M 2002 \JHEP~ {\bf 0205} 037 \\
Chen Y X,   Hou B Y, Hou B Y 2002 \NPB~ {\bf 638} 220 \\
Kimura Y 2002  \NPB~ {\bf 637} 177 \\
Bernevig B A, Chern C H, Hu J P, Toumbas N and Zhang S C 2002
\APNY {\bf 300} 185 \\
Zhang S C 2003 \PRL~ {\bf 90} 196801 \\
Dolan B 2003 \JHEP~ {\bf 0305} 18 \\ 
Meng G 2003 \IJMPA~ {\bf 36} 9415 \\
Bellucci S, Casteill P Y and  Nersessian A 2003 \PLB~ {\bf 574} 121\\
Hasebe K and Kimura Y 2004 \PLB~ {\bf 602} 255

\bibitem{eff-action} Elvang H and Polchinski J 2002 {\it Preprint} hep-th/0209104
 
\bibitem {NR1} Nair V P and Randjbar-Daemi S 2004 \NPB~
{\bf 679} 447

\bibitem{S8} Bernevig B A, Hu J P, Toumbas N and Zhang S C 2003 \PRL~
{\bf 91} 236803 

\bibitem{jellal}  Jellal A 2005 \NPB~ {\bf 725} 554\\
Daoud M and Jellal A 2006 {\it Preprint} hep-th/0605289

\bibitem{poly3} Polychronakos A P 2005 \NPB~ {\bf 705}  457 \\
Polychronakos A P 2005 \NPB~ {\bf 711}  505

\bibitem{haldane} Haldane F D M 1983 \PRL~ {\bf 51}  605

\bibitem{KA2} Karabali D 2005 \NPB~ {\bf 726} 407 \\
Karabali D 2006 {\it Preprint} hep-th/0605006


\bibitem{IKS} Iso S, Karabali D and Sakita B 1992 \PLB { \bf 296} 143 \\
Cappelli A, Dunne G, Trugenberger C and Zemba G 1993 \NPB~ {\bf 398} 531 \\
Cappelli A, Trugenberger C and Zemba G 1993 \NPB~ {\bf 396} 465 \\
Cappelli A, Trugenberger C and Zemba G 1994 \PRL~ {\bf 72} 1902 \\
Karabali D 1994 \NPB~ {\bf 419} 437 \\
Karabali D 1994 \NPB~ {\bf 428} 531 \\
Flohr M and Varnhagen R 1994 \JPA~ {\bf 27} 3999

\bibitem{wadia} Das S R, Dhar A, Mandal G and Wadia S R 1992 
{\it Int. J. Mod. Phys. A} {\bf 7} 5165 \\   
Das S R, Dhar A, Mandal G and Wadia S R 1992 \MPLA~ {\bf 7} 71 \\
 Dhar A, Mandal G and Wadia S R 1993  \IJMPA~ {\bf 8}  325 \\ 
 Dhar A, Mandal G and Wadia S R 1992 \MPLA~ {\bf 7}  3129 \\
 Dhar A, Mandal G and Wadia S R 1993 \MPLA~ {\bf 8} 3557 \\
 Dhar A 2005 \JHEP~ {\bf 0507}  064
 
 \bibitem{sakita} Sakita S 1993 \PLB~{\bf 315}  124\\
 Sakita B 1996 \PLB~ {\bf 387} 118 \\
Sakita B and Ray R 2001 \PRB~ {\bf 65}  035320

\bibitem{shizuya} A similar approach, also for two-dimensional QHE, but
based on $W_{\infty}$-transformations mixing higher Landau levels has been proposed by  Shizuya K 1995 \PRB~ {\bf 52}  2747

\bibitem{seib1} Seiberg N and Witten E 1999 \JHEP~ {\bf 9909}  032

\bibitem{seib2} Madore J, Schraml S, Schupp P and Wess J 2000 {\it Eur. Phys. J.}  C {\bf 16} 161 \\
Jurco B, Moller L , Schraml S, Schupp P and Wess J 2001 {\it Eur. Phys. J.} C {\bf 21} 383 \\
Behr W and Sykora A 2004  \NPB~ {\bf 698}  473

\bibitem{bal5} Alexanian G, Balachandran A P, Immirzi G and  Ydri B 2002 \JGP~ {\bf42} 28 \\ Balachandran A P, Dolan B P, Lee J, Martin X and O'Connor D 2002 \JGP~ {\bf 43} 184

\bibitem{nepomechie} Brown L S and Nepomechie R I 1987  \PRD~ {\bf 35}  3239 \\
 Karabali D, Park Q H, Schnitzer H J and Yang Z 1989  \PLB~ {\bf 216} 307 \\
 Karabali D and Schnitzer H J 1990  \NPB~ {\bf 329} 649

\bibitem{NS} Nair V P and Schiff J 1990 \PLB~ {\bf 246}  423 \\
Nair V P and Schiff J 1992 \NPB~ {\bf 371} 329 \\ see also,
Donaldson S 1985 {\it Proc. London Math. Soc.} (3) {\bf 50} 1 \\
Losev A {\it et al}  1996 {\NPBProc} {\bf 46}  130 \\
Inami T,  Kanno H and  Ueno T 1997  \MPLA ~
{\bf 12}  2757 

\bibitem{wen} Wen X G 1990 \PRB~ {\bf 41} 12838 \\
Lee D H and Wen X G  1991 \PRL~ {\bf 66}  1765 \\
Stone M 1990 \PRB~ {\bf 42}  8399 \\
Stone M 1991 \APNY {\bf 207}  38 \\
Frohlich J and Kepler T 1991 \NPB~ {\bf 354} 369

\bibitem{KNP} Karabali D, Nair V P and Polychronakos A P 2002
\NPB~ {\bf 627}  565 \\
Grosse H and Steinacker H 2005
\NPB~ {\bf 707} 145

\bibitem{bal3}  Nair V P and  Randjbar-Daemi S 1998
\NPB ~{\bf 533} 333 \\
Grosse H and  Strohmaier A 1999 {\it Lett. Math.
Phys.} ~{\bf 48} 163 \\
Alexanian G ,  Pinzul A and Stern A 2001 \NPB ~{\bf
600} 531

\bibitem{nair1} Nair V P 2006 {\it Preprint} hep-th/0605007

\bibitem{nair2} Nair V P 2006 {\it Preprint} hep-th/0605008

\bibitem{abe} Nair V P 2003  \NPB~{\bf 651} 313\\
Abe Y and Nair V P 2003
\PRD~{\bf 68} 025002

\bibitem{zanelli} Chamseddine A H 1989 \PLB ~{\bf 233}  291 \\
Chamseddine A H  1990 \NPB~ {\bf 346} 213 \\
 Witten E 1988 \NPB~{\bf 311}  46; \\
 Achucarro A and Townsend P 1986  \PLB~{180} 89 \\
For a recent review on Chern-Simons gravity, see,
Zanelli J 2000 {\it Braz. J. Phys.} {\bf 30}  251 and Zanelli J 2005 {\it Preprint}  hep-th/0502193

\bibitem{horava} Troncoso R and Zanelli J 1998 \PRD~ {\bf 58} 101703 \\
Horava P 1999 \PRD ~{\bf 59} 046004 \\
Nastase H 2003 {\it Preprint} hep-th/0306269 \\
Horava P and Keeler C A 2005 {\it Preprint} hep-th/0508024

 \bibitem{CMM} This is related to the description of gravity developed in
 Mansouri F and Chang L N 1976 \PRD~
 {\bf 13} 3192\\
 MacDowell S and Mansouri F 1977 \PRL~
 {\bf 38} 739\\
Mansouri F 1977 \PRD~ {\bf 16} 2456

\bibitem{poly1} Chamseddine A H and Frohlich J 1994
\JMP ~{\bf 35} 5195 \\
Krajewski T 1998 {\it Preprint} math-ph/9810015 \\
Mukhi S and Suryanarayana N V 2000 \JHEP~{\bf 0011} 006 \\
Polychronakos A P 2000 \JHEP~{\bf 0011} 008 \\
Chen G H and Wu Y S 2001 \NPB~{\bf 93}  562 \\
Grandi N V and  Silva G A 2001 \PLB ~{\bf 507}  345

\bibitem{poly2} Morariu B and Polychronakos A P 2005
\PRD ~{\bf 72} 125002

\bibitem{boso1} Luther A 1979 \PRB~ {\bf 19}  320 \\
Haldane F D 1992  {\it Helv. Phys. Acta} {\bf 65} 152 ({\it Preprint} cond-mat/0505529) \\
Castro Neto A H and Fradkin E 1994 \PRB~ {\bf 49}  10877 \\
Houghton A and Marston B 1993 \PRB~ {\bf 48}  7790 \\
Kwon H J, Houghton A and Marston B 1995 \PRB~ {\bf 52} 8002 \\
Anderson P W and Khveshchenko D 1995 \PRB~{\bf 52}  16415 \\
Dhar A, Mandal G and Suryanarayana N V 2006 \JHEP~ {\bf 0601} 118 \\
Dhar A and Mandal G 2006 {\it Preprint} hep-th/0603154

\bibitem{poly4} Polychronakos A P 2005 {\it Preprint} hep-th/0502150 
 

\bibitem{fluids} Jackiw R and Polychronakos A P 1999 \CMP~ {\bf 207} 107\\
Jackiw R, Nair V P, Pi S-Y and Polychronakos, A P 2004 \JPA {\bf 37} R327

\bibitem{suss}
Susskind L 2001 {\it Preprint} hep-th/0101029\\
Polychronakos A P 2001 \JHEP~{\bf 0104} 011

\bibitem{beren} See, for example, Berenstein D 2004 \JHEP~{\bf 0407}  018 \\
Lin H, Lunin O and Maldacena J 2004 \JHEP~{\bf 0410}  025 \\
Bernevig B A, Brodie J, Susskind L and Toumbas N 2001
\JHEP~{\bf 0102} 003 \\
Gubser S and  Rangamani M 2001 \JHEP~ {\bf 0105}  141 \\
Castro C 2004 \GRG~{\bf 36} 2605

\bibitem{hasebe}
Hasebe K 2005 \PRL~{\bf 94} 206802\\
Hasebe K 2005 {\it Preprint} hep-th/0503162\\
Hasebe K 2006 {\it Preprint} hep-th/0606007

%%%%%%%%%%%%%%%%%%%%%%%%%%%%%%%%%%%%%%%%%%%%%%

\end{thebibliography}
\end{document}